\documentclass[sigconf,plain,authorversion,manuscript]{acmart}

\renewcommand\footnotetextcopyrightpermission[1]{} 
\AtBeginDocument{%
  \providecommand\BibTeX{{%
    \normalfont B\kern-0.5em{\scshape i\kern-0.25em b}\kern-0.8em\TeX}}}

\usepackage{multirow}
\usepackage{multicol}
\usepackage{algorithm2e}
\usepackage{listings}
\usepackage{todonotes}
\usepackage{tabularx}

\usepackage[symbol]{footmisc}

\definecolor{codegreen}{rgb}{0,0.6,0}
\definecolor{codegray}{rgb}{0.5,0.5,0.5}
\definecolor{codepurple}{rgb}{0.58,0,0.82}
\definecolor{backcolour}{rgb}{0.95,0.95,0.92}

\lstdefinestyle{mystyle}{
    backgroundcolor=\color{backcolour},
    commentstyle=\color{codegreen},
    keywordstyle=\color{magenta},
    numberstyle=\tiny\color{codegray},
    stringstyle=\color{codepurple},
    basicstyle=\ttfamily\footnotesize,
    breakatwhitespace=false,
    breaklines=true,
    captionpos=b,
    keepspaces=true,
    numbers=left,
    numbersep=5pt,
    showspaces=false,
    showstringspaces=false,
    showtabs=false,
    tabsize=2
}

\begin{document}

\title{A Portable Framework for Accelerating Stencil Computations on Modern Node Architectures}

\author{Ryuichi Sai}
\email{ryuichi@rice.edu}
\affiliation{%
  \institution{Rice University}
  \streetaddress{6100 Main Street, MS-132}
  \city{Houston}
  \state{Texas}
  \country{USA}
  \postcode{77005}
}

\author{John Mellor-Crummey}
\affiliation{%
  \institution{Rice University}
  \streetaddress{6100 Main Street, MS-132}
  \city{Houston}
  \state{Texas}
  \country{USA}
  \postcode{77005}
}

\author{Jinfan Xu}
\affiliation{%
  \institution{Rice University}
  \streetaddress{6100 Main Street, MS-132}
  \city{Houston}
  \state{Texas}
  \country{USA}
  \postcode{77005}
}

\author{Mauricio Araya-Polo}
\affiliation{%
  \institution{TotalEnergies EP Research \& Technology US, LLC.}
  \city{Houston}
  \state{Texas}
  \country{USA}
}

\renewcommand{\shortauthors}{Sai, et al.}

\begin{abstract}
\sloppy

Finite-difference methods based on high-order stencils are widely used in seismic simulations, weather forecasting, computational fluid dynamics, and other scientific applications. Achieving HPC-level stencil computations on one architecture is challenging, porting to other architectures without sacrificing performance requires significant effort, especially in this golden age of many distinctive architectures.

To help developers achieve performance, portability, and productivity with stencil computations, we developed StencilPy. With \hbox{StencilPy}, developers write stencil computations in a high-level domain-specific language, which promotes productivity, while its backends generate efficient code for existing and emerging architectures, including modern many-core CPUs (such as AMD Genoa-X, Fujitsu A64FX, and Intel Sapphire Rapids), latest generations of GPUs (including NVIDIA H100 and A100, AMD MI200, and Intel Ponte Vecchio), and accelerators (including  Cerebras and STX). StencilPy demonstrates promising performance results on par with hand-written code, maintains cross-architectural performance portability, and enhances productivity. Its modular design enables easy configuration, customization, and extension.

A 25-point star-shaped stencil written in StencilPy is one-quarter of the length of a hand-crafted CUDA code and achieves similar performance on an NVIDIA H100 GPU. In addition, the same kernel written using our tool is $7\times$ shorter than hand-optimized code written in Cerebras Software Language (CSL), and it delivers comparable performance that code on a Cerebras CS-2.

\end{abstract}





\maketitle

\fancyfoot{}
\thispagestyle{empty}

\section{Introduction}

With the diversity of architectures available to application developers, achieving performance, portability, and productivity (3Ps) has become increasingly important. Adopting a 3Ps approach can simplify the development and maintenance of scientific applications, enabling them to run efficiently on multiple architectures without sacrificing performance.

However, achieving this goal has always been challenging due to the inherent trade-offs among performance, portability, and productivity. Achieving high performance and portability requires implementing algorithms that exploit low-level hardware features. On the other hand, productivity is accomplished with higher-level software abstractions that hide these low-level details.
The problem becomes more significant with modern node architectures, as low-level hardware details vary between architectures, vendors, and even different generations of products from the same vendor.
Additionally, hardware vendors lack unified programming models, and occasionally, even the programming models and frameworks provided by the same vendor change significantly over time.
As a result, developing, optimizing, and maintaining scientific applications is costly due to the diversity of architectures, programming models, and the evolution of vendor software stacks.

Stencil computation is widely used in many scientific applications, such as modeling of seismic wave propagation, weather forecasting, computational fluid dynamics, and convolutional neural networks.
Therefore, achieving 3Ps for stencil computations is of great interest.
While a 25-point star-shaped stencil update from a global view requires only ten lines of C code,
our highly optimized GPU kernel for the same stencil computation required 375 lines of CUDA code and more than 600 lines of auxiliary code in C~\cite{ryuichi_pmbs_2020,ryuichi_cpe_2021}.
Furthermore, manually porting a similar stencil code to a Cerebras CS-2 system (CS2) needed 1613 lines of CSL code and 364 lines of Python code~\cite{cerebras_sdk_doc}.
Accelerating such kernels on these platforms requires complex algorithms, optimization strategies, careful data allocations, and handling intricate data movement.
Therefore, they are also difficult to program and unable to port between systems.

A portable framework for accelerating stencil computations is needed to address these challenges and achieve 3Ps for high-order stencil computations on modern node architectures.
To address this need, we have been developing StencilPy---a portable Python framework for implementing stencil computations.
The framework's frontend enables developers to express common stencil computations using a high-level abstraction with a global-view logic, thus maintaining high developer productivity.
Its platform-agnostic syntax and kernel launch invocation facilitate porting to various systems without changing the implementation.
The backend generates high-performance executables for modern node architectures, including modern GPUs and several accelerators.
The framework also explores ways to reduce overhead during code analysis and generation while providing a user-friendly interface for domain experts to specify optimizations and facilitate evaluations.

This manuscript presents our progress on StencilPy, a portable framework for accelerating stencil computations on modern node architectures, including a Python-hosted domain-specific language and code generators for
modern many-core CPUs (such as AMD Genoa-X, Fujitsu A64FX~\cite{fujitsu_a64fx}, and Intel Sapphire Rapids), latest generations of GPUs (including NVIDIA H100 and A100, AMD MI200, and Intel Ponte Vecchio), and accelerators (including  Cerebras~\cite{cerebras_hotchips2019} and STX~\cite{epi_stx_hotchips}).
While our goal is to support stencils of commonly used shapes and different orders, our effort to date has focused primarily on high-order stencils. In this chapter, we evaluate our  framework using a 25-point star-shaped stencil commonly used by the energy industry for the acoustic isotropic approximation of the wave equation as part of seismic imaging ~\cite{meng_minimod_2020}.
This manuscript describes the following contributions:
\begin{itemize}
    \item the design of an embedded domain-specific language (DSL) for expressing stencil computations in Python;
    \item the implementation of a framework that parses and analyzes the DSL, generates platform-specific code for modern CPUs (AMD Genoa-X, Apple Silicons, Fujitsu A64FX, IBM Power, and Intel Sapphire Rapids), multiple generations of GPUs of different vendors (NVIDIA H100, A100, and V100; AMD MI200 and MI100; and Intel Ponte Vecchio), and accelerators (STX~\cite{epi_stx_hotchips} and Cerebras CS-2~\cite{cerebras_hotchips2019});
    \item optimization strategies that minimize overheads associated with the StencilPy framework and accelerate runtime performance; and
    \item a performance evaluation based on a 25-point star-shaped high-order stencil used for seismic imaging.
\end{itemize}

The next section provides some background about stencil computations.
Section~\ref{sec:related} reviews related work.
Section~\ref{sec:stencilpy-arch} describes the framework's design and architecture.
Section~\ref{sec:stencilpy-dsl-fe} introduces the DSL and frontend.
Section~\ref{sec:stencilpy-franework-impl} describes the implementation of the StencilPy framework.
Section~\ref{sec:stencilpy-opt} provides an overview of StencilPy's optimizations to boost stencil runtime performance.
Sections~\ref{sec:stencilpy-workflow}~and~\ref{sec:stencilpy-backend-templates} describe the framework's workflow and its backend templates, respectively.
Section~\ref{sec:stencilpy-ir} describes StencilPy's intermediate representations.
Section~\ref{sec:stencilpy-codegen} extensively describes code generation.
Section~\ref{sec:stencilpy-cus-ext} explores options to customize and extend the framework beyond its current capabilities.
Section~\ref{sec:stencilpy-eval} describes our evaluation methodologies to assess numerical correctness, performance, portability, and productivity, and then it reviews our findings.
Sections~\ref{sec:stencilpy-summary}~and~\ref{sec:stencilpy-future-work} summarize our conclusions and discuss future work.

\section{Background}
\label{sec:background}

To help the reader understand the domain applications StencilPy targets, we provide a brief introduction to high-order stencil computations, especially ones used in seismic modeling.

\subsection{High-Order Stencil Computations}

In stencil computations, data elements from a multi-dimensional array are iteratively updated according to a fixed pattern. In this work, an array is manipulated as a Cartesian grid. An element in the grid is usually called a cell or a point. Calculating the next value for a cell using a stencil involves computing a weighted sum of products between values of a set of neighboring cells (stencil defines the set of cells used) and scaling coefficients.

Applying a stencil pattern to the points in a block requires values for points in neighboring blocks. The points needed from neighboring blocks are collectively known as the halo region. The thickness of the halo along each dimension is called the halo size or halo width, and it also defines the order of the stencil. When a stencil has a large halo width, it is called a high-order stencil. Stencil computations are applied to all the elements of a grid over a sequence of iterations, until the underlying equation coverges.

\subsection{Seismic Modeling and Acoustic Isotropic Approximation}

In this manuscript, we study high-order stencil-based implementations of the acoustic isotropic approximation (Acoustic ISO) of the wave equation~\cite{meng_minimod_2020},
which is commonly used by the energy industry on large grids to model and characterize the subsurface for a variety of purposes.

The wave equation  for an acoustic isotropic operator with constant-density has the following form:
\begin{equation}
\frac{1}{\mathbf{V}^2}\frac{\partial^2 \mathbf{u}}{\partial t^2} - \nabla^2 \mathbf{u} = \mathbf{f},
\end{equation}
where $\mathbf{u} = \mathbf{u}(x,y,z)$ is the wavefield, $\mathbf{V}$ is the Earth model (with velocity as rock property), and $\mathbf{f}$ is the source perturbation. The equation is discretized in time using a second-order centered stencil, resulting in the semi-discretized equation:
\begin{equation}
\mathbf{u}^{n+1} - \mathbf{Q}\mathbf{u}^n + \mathbf{u}^{n-1} = \left(\Delta t^2\right) \mathbf{V}^2 \mathbf{f}^n,\label{eq:minimod-semidisc}
	\mathrm{with\ }\mathbf{Q} = 2 + \Delta t^2 \mathbf{V}^2 \nabla^2.
\end{equation}
Finally, the equation is discretized in space using a 25-point stencil in 3D, with eight points in along each axis surrounding a center point,
where $c_{xyz}, c_{xm}, c_{ym}, c_{zm}$ are the discretization parameters:

\begin{multline}
\mspace{200mu} \nabla^2 \mathbf{u}(x,y,z) \approx c_{xyz} \times \mathbf{u}(i,j,k) + \\
\mspace{45mu} \sum_{m=1}^4 (c_{xm}\times\left[\mathbf{u}(i+m,j,k) + \mathbf{u}(i-m,j,k)\right] + \\
\mspace{75mu} c_{ym}\times\left[\mathbf{u}(i,j+m,k) + \mathbf{u}(i,j-m,k)\right] + \\
\mspace{75mu} c_{zm}\times\left[\mathbf{u}(i,j,k+m) + \mathbf{u}(i,j,k-m)\right]) \\
\end{multline}

A high-level description of the algorithm is shown in Algorithm \ref{algo:minimod}.
As is common for seismic modeling, the simulations employ a Perfectly-Matched Layer (PML) \cite{komatitsch-pml} boundary condition around the simulation domain. The resulting extended domain consists of an ``inner'' region and a surrounding ``PML'' region.
As described later in this manuscript, our framework is designed to decompose the data domain and launch dedicated kernels accordingly to achieve great performance.

\begin{algorithm}[t]
	\KwIn{\\\Indp \Indp
    $\mathbf{f}$: source}
	\KwOut{\\\Indp \Indp
    $\mathbf{u}^n$: wavefield at timestep $n$, for $n\leftarrow 1$ \KwTo $T$}
	$\mathbf{u}^0 := 0$\;
	\For{$n\leftarrow 1$ \KwTo $T$}{\nllabel{line:tsloop}
		\For{each point in wavefield $\mathbf{u}^n$}{
			Solve Eq.~\ref{eq:minimod-semidisc} (left hand side) for wavefield $\mathbf{u}^n$\;
		}
		$\mathbf{u}^n = \mathbf{u}^n + \mathbf{f}^n$ (Eq.~\ref{eq:minimod-semidisc} right hand side)\;
	}\nllabel{line:end-of-ts}
	\caption{A high-level description of the algorithm for solving the acoustic isotropic approximation of the wave equation with constant density.}
	\label{algo:minimod}
\end{algorithm}

\begin{figure}
\centering
\includegraphics[width=0.42\columnwidth]{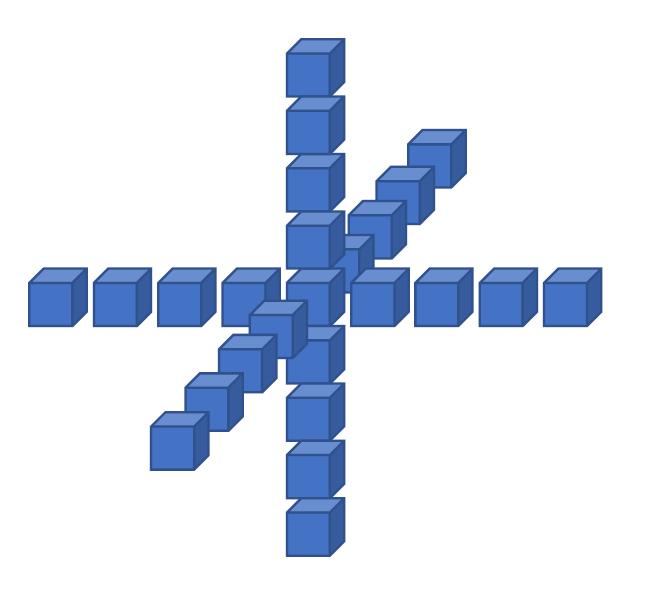}
\captionof{figure}{A star-shaped 25-point stencil.}
\label{figure:stencil-star-25}
\end{figure}

This kernel involves applying a star-shaped 25-point stencil, as depicted in Figure~\ref{figure:stencil-star-25}, to elements of a 3D array.
Computation in the PML region is more complex than that in the inner region.
The PML region employs the same 25-point stencil applied in the inner region and also a 7-point star-shaped stencil to a different array to compute boundary conditions.

The grid, representing the physical domain, tends to be substantial in production simulations. Each of its dimensions is usually large with up to thousands of points. To simulate how the waves propagate through the domain, it is necessary to apply the stencil computations iteratively for a large number of time steps.

\section{Related Work}\label{sec:related}

There is much prior work~\cite{frigo_cache-oblivious_1999,frigo_cache_2005,frigo_cache_2006,strzodka_cache_2010,tang_pochoir_2011,wonnacott_using_2000,wonnacott_achieving_2002,jin_increasing_2001,mccalpin_time_1998,song_new_1999,de_la_cruz_introducing_2010,de_la_cruz_algorithm_2014,krishnamoorthy_effective_2007, holewinski_high-performance_2012,grosser_split_2013,nguyen_35-d_2010,micikevicius_3d_2009,matsumura_an5d_2020,rawat_sdslc_2015,rawat_domain-specific_2018} exploring accelerating stencil computations using a compiler-assist approach.
Here, we describe the ones most relevant to our work.

\paragraph{Stencil Optimizations}

Time skewing~\cite{wonnacott_using_2000,wonnacott_achieving_2002,jin_increasing_2001,mccalpin_time_1998,song_new_1999} avoids costly data movement by skewing data dimension(s) by the time dimension so that cached data is reused for multiple time steps.
Cache-oblivious algorithms~\cite{frigo_cache-oblivious_1999,frigo_cache_2005,frigo_cache_2006,strzodka_cache_2010,tang_pochoir_2011} tile the domain and performs a space cut or a time cut to maximize the use of each memory level.
Overlapped tiling uses time skewing to trade redundant computation along the boundaries of overlapped tiles for a reduction in memory bandwidth required \cite{krishnamoorthy_effective_2007,holewinski_high-performance_2012}.
Split tiling \cite{grosser_split_2013}, on the other hand, mitigates redundant computations by employing a two-phase computation approach. In the first phase, hyper-trapezoidal tiles along the time dimension are used, and in the second phase, the missing points are backfilled.

Streaming on the outermost dimension is a widely used optimization technique for accelerating high-order stencils on GPUs~\cite{ryuichi_cpe_2021,ryuichi_pmbs_2021}.
Nguyen et al.~\cite{nguyen_35-d_2010} introduced a 2.5D spatial blocking technique, which was further extended to a 3.5D blocking algorithm by combining it with 1D temporal blocking.
Micikevicius~\cite{micikevicius_3d_2009} used registers to store data elements along a streaming dimension to improve performance.
Matsumura et al.~\cite{matsumura_an5d_2020} refined this approach by incorporating fixed register allocations, double buffering, and a division of the streaming dimension.
In our work, we leverage these optimization techniques for GPU and STX backends in our code generation and optimization process.

The Semi-stencil algorithm~\cite{de_la_cruz_introducing_2010,de_la_cruz_algorithm_2014} is a technique that divides a stencil computation into two parts: a forward update and a backward update. This partitioning offers a tradeoff between the number of loads and stores involved in the computation.
In our work, we leverage the Semi-stencil algorithm for certain high-order stencils on both GPU and STX backends. This approach proves advantageous due to the significant reduction in the number of loads required for high-order stencils, resulting in improved performance.

While previous work primarily emphasizes performance and relies on hand-crafted code for evaluation, our work also highlights performance portability and productivity by leveraging the power of compiler technologies.

\paragraph{Stencil DSLs and automated code generations}

Domain-Specific Languages (DSLs) for stencil computations have been widely studied~\cite{tang_pochoir_2011,rawat_domain-specific_2018,rawat_register_2018,rawat_optimizing_2019,baghdadi_pencil_2015,baghdadi_tiramisu_2019,christen_patus_2011,louboutin_devito_2019,ragan-kelley_halide_2013}.
Rawat et al.~\cite{rawat_register_2018}  use Directed Acyclic Graphs (DAGs) to analyze register dependencies and optimize the ordering of registers for improved performance. Rawat et al.~\cite{rawat_optimizing_2019} later enhance this approach by incorporating dynamic resource allocations with automatic tuning techniques.
BrickLib~\cite{bricklib_p3hpc_2023} leverages fine-grained data blocking and generates vector code for the blocks to achieve great performance portability on CPUs and GPUs.
Diamond tiling using a polyhedral model~\cite{bandishti_tiling_2012,bondhugula_practical_2008} has been studied and integrated into modern compilers and their toolchains.
Several prior work~\cite{fuhrer_towards_2014,gorius_modeling_2019,gysi_domain-specific_2020} propose various reusable designs for multi-layer intermediate representations, aiming to enhance the utilization of these representations between DSL frontends and code generations.
STELLA~\cite{gysi_stella_2015} integrates within the standard C++ template meta-programming, and application developers can fall back on the host language
if they need to incorporate features not supported by the library.
Halide~\cite{ragan-kelley_halide_2013} introduces a DSL embedded in C++ rather than being a standalone language. It builds an in-memory representation using Halide's C++ API, which can be compiled ahead-of-time or just-in-time.

For distributed memory architectures, such as the ones in dataflow architecture, stencil computation strategies are studied in High Performance Fortran~\cite{roth_stencil_compilation_strategy,roth_hpf_compiling_stencils}.

Functional programming~\cite{steuwer_lift_2017,lucke_functional_2020} has also been explored in the past for robust reasoning of the stencil computations and their optimizations.

These studies inspire our work, and we also aim for performance portability and developer productivity.
Our framework also generates code for a broader range of modern node architectures.

\paragraph{Software Frameworks in Python}

We have drawn inspiration from Numpy~\cite{numpyHarris2020array}---a comprehensive Python library for scientific numerical computations.
Numpy implements performance-critical components in the C language while keeping the user-facing frontend in Python.
We adopt a similar approach, but our specialization is stencil computations rather than general-purpose computations.

Numba~\cite{lam2015numba} delegates machine code optimizations and generations at runtime to LLVM~\cite{lattner_llvm_2004}. It leverages the standard Python interpreter while annotating Numba code regions with decorators.
Similarly, we use decorators, but our approach involves generating code using platform-specific programming models and employing platform-specific compilers for building the generated code.

Other specialized frameworks, such as Taichi~\cite{taichi}, TensorFlow~\cite{tensorflow2015-whitepaper}, and PyTorch~\cite{pytorch}, follow a similar architectural design to Numpy.
Mojo~\cite{modular_mojo} is a superset of Python that uses LLVM~\cite{lattner_llvm_2004} and MLIR~\cite{lattner_mlir_2020} for code generation. It uses inferred static typing.
However, these frameworks primarily focus on domains such as machine learning, image processing, or artificial intelligence, whereas our framework specializes in stencil computations.
While our framework does not support type inference, our code optimization and generation need type information. Therefore, our framework requires type hints for stencil target and kernel functions.

APPy~\cite{vivek_2024} annotates Python code using a syntax similar to that of OpenMP's \texttt{pragma}s. It focuses on loop-level optimization and automates code vectorization on GPUs.  Similarly, our framework offers GPU templates that facilitate vectorization, with a particular focus on Stencil computation.

Frameworks specifically designed for stencils, such as PyStencil~\cite{pystencil} and Devito~\cite{luporini_architecture_2020,louboutin_devito_2019}, offer domain application developers the ability to express their applications at high levels of abstraction, such as in the domains of physics and mathematics. These frameworks automatically apply stencil computations while transforming high-level syntax to machine code and incorporating various optimizations.
In contrast, our framework provides an interface that operates at the level of computational abstraction. This approach maximizes productivity by supporting expressive stencil computations without sacrificing optimization opportunities and maintaining code simplicity.
Moreover, our framework supports broader backends, including emerging architectures.




\section{Framework Design and Architecture}
\label{sec:stencilpy-arch}

\begin{figure*}
    \centering
    \includegraphics[width=0.9\textwidth]{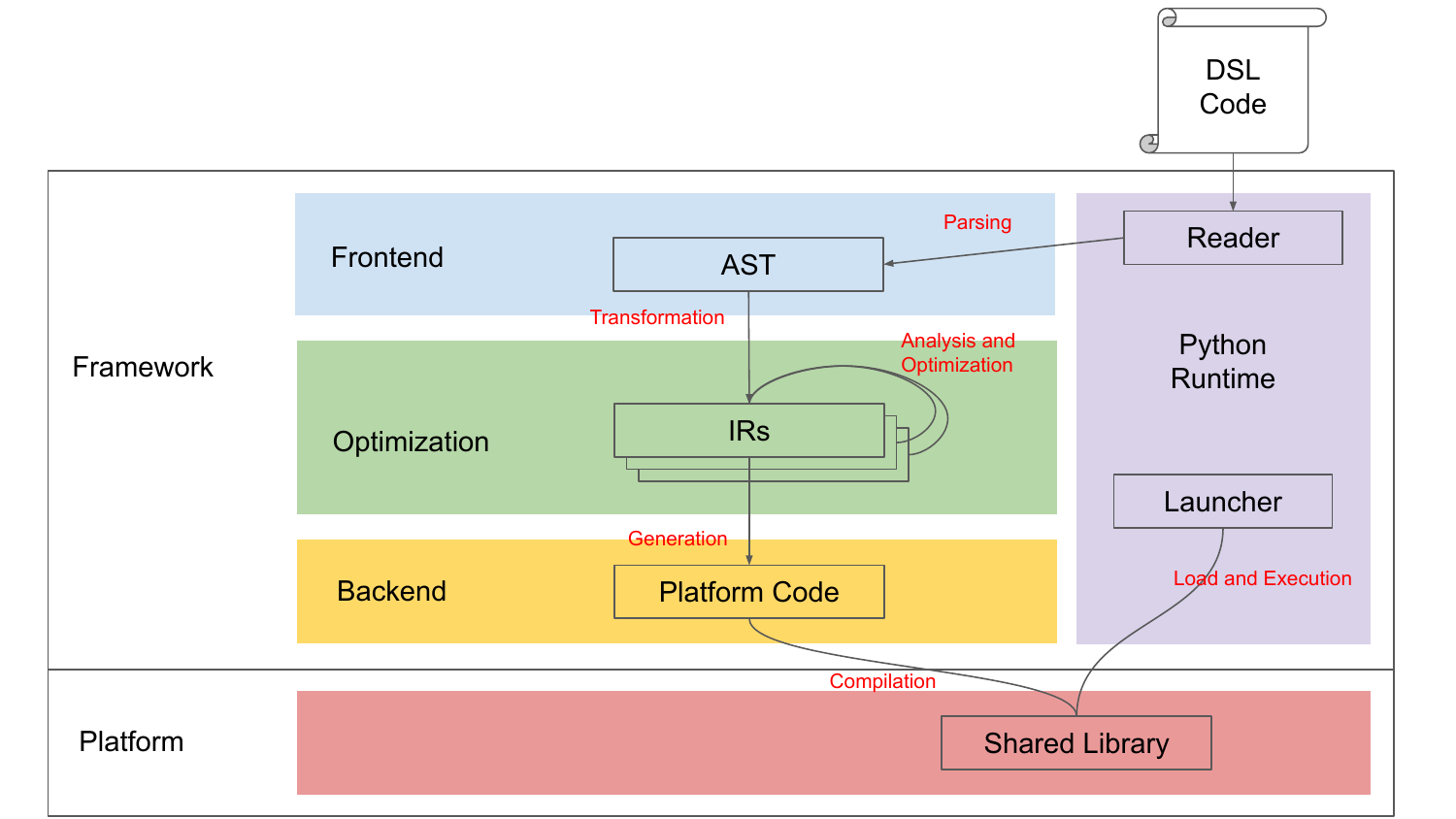}
    \caption{The StencilPy framework architecture.}
    \label{figure:stencilpy-framework-architecture}
\end{figure*}

The StencilPy framework consists of multiple layers, as illustrated in Figure~\ref{figure:stencilpy-framework-architecture}.

The frontend layer enables application developers to express stencil computations using a DSL embedded in Python. The frontend parses the DSL code and converts it into an abstract syntax tree representation.

The optimization layer serves several purposes. First, it applies static and semantic analyses to understand the stencil shape, computation pattern, and data characteristics.
The optimization layer employs multiple passes, including decomposition, tiling, reordering, and others to transform the stencil computations by lowering a high-level abstraction to a form optimized for execution on the target backend.
Multiple intermediate representations are designed and used during the process to support these transformations.

The backend is capable of producing high-performance code for a range of modern architectures and generates a corresponding code version based on configuration.

Last, the launcher employs Just-In-Time compilation of the code, executes the program, and captures the results.

\section{DSL and Frontend Design}
\label{sec:stencilpy-dsl-fe}

Listing~\ref{lst:stencilpy-star2d4r} depicts an example of a star2d4r stencil implementation using the StencilPy framework.

\begin{lstlisting}[float,floatplacement=t,language=Python, basicstyle=\linespread{1.2}\ttfamily\scriptsize, label={lst:stencilpy-star2d4r}, caption=A star2d4r stencil implementation using StencilPy.]
import stencilpy as st

@st.kernel
def kernel_star2d4r(u: st.grid, v: st.grid):
  v.at(0, 0).set(0.25005 * u.at(0, 0)
    + 0.11111 * (u.at(-4, 0) + u.at(4, 0))
    + 0.06251 * (u.at(-3, 0) + u.at(3, 0))
    + 0.06255 * (u.at(-2, 0) + u.at(2, 0))
    + 0.06245 * (u.at(-1, 0) + u.at(1, 0))
    + 0.06248 * (u.at(0, -1) + u.at(0, 1))
    + 0.06243 * (u.at(0, -2) + u.at(0, 2))
    + 0.06253 * (u.at(0, -3) + u.at(0, 3))
    - 0.22220 * (u.at(0, -4) + u.at(0, 4)))

@st.target
def target_star2d4r(u: st.grid, v: st.grid, iter:st.i32):
  for _t in range(iter):
    st.map(e=u.shape)(kernel_star2d4r)(u, v)
    (v, u) = (u, v)

u = st.grid(dtype=st.f32, shape=(1000,1000), order=4)
v = st.grid(dtype=st.f32, shape=(1000,1000), order=4)
# data initialization omitted for brevity
st.launch(
    backend=st.cuda(
        computeCapability="9.0",
        threadsPerBlock=(16, 8, 8),
        template=st.CUDABackend.Template.gmem,
    )
)(target_star2d4r)(u, v, 1000)
\end{lstlisting}

StencilPy's DSL is hosted in Python by expanding its syntax.
Domain scientists commonly use Python because of its simple, high-level syntax. Python is ideal for rapid prototyping.
It has an extensive library ecosystem that further enhances its utility.
Python's user-friendly nature contributes to its ease of learning and smooth transition from implementations in other languages.
Our DSL is embedded in Python to leverage these pre-existing advantages.

Our DSL distinguishes itself from Python in two areas:
1) we introduce a few constructs specific for stencil computations, and
2) in our constructs, type hints are required.

\paragraph{StencilPy-Specific Constructs}

Table~\ref{table:stencilpy-dsl-constructs} shows all constructs and their purposes introduced by the StencilPy framework.

\begin{table*}[t]
\centering
\begin{tabular}{|c|p{12.9cm}|}
\hline
Construct & \multicolumn{1}{c|}{Purpose}                                                                                                        \\ \hline
kernel    & A compute kernel on device. It is typically the stencil loop.                                                                                       \\ \hline
target    & Host logic to set up the computations and launch kernels. Usually, this is the time loop for stencil iterations.                   \\ \hline
map       & Looping the multi-dimensions in a data grid, and maps a kernel to each stencil point.                                                 \\ \hline
launch    & Specifies the backend used in the stencil computation, optimization strategies employed in the simulation, and launches the target. \\ \hline
at        & Reads a point value based on the offset indices from the current stencil point.                                                     \\ \hline
at.set    & Updates a point value based on the offset indices.                                                                                  \\ \hline
grid      & A data array used to store a stencil data grid.                                                                                     \\ \hline
\end{tabular}
    \caption{StencilPy DSL constructs.}
    \label{table:stencilpy-dsl-constructs}
\end{table*}

By annotating \texttt{@st.target} or \texttt{@st.kernel}, the regular Python function declarations become StencilPy targets or kernels, respectively.

Each \texttt{kernel} implicitly defines the indices of the current center point, which is used as a base to offset from in \texttt{at} and \texttt{at.set} constructs.

A \texttt{map} construct takes three argument groups: looping pattern, pointers to the kernels, and kernel parameters.
When Perfectly-Matched Layers (PML)~\cite{komatitsch-pml} present, to support inner region and PML regions, StencilPy takes the looping pattern explicitly defined with the begin and end indices of the inner region and PML regions.
In addition, to improve developer expressiveness, the framework provides syntactic sugar for common use cases where the region boundaries can be inferred from the stencil data grid.
For example, the \texttt{map} call in Listing~\ref{lst:stencilpy-star2d4r} shows a pattern that loops through the whole grid determined by the shape of the $u$ grid.

To support various backends, when \texttt{launch}-ing a target, backend-specific parameters are provided in a \texttt{backend} object.
Listing~\ref{lst:stencilpy-star2d4r} shows an example of launching a backend leveraging the CUDA programming model.
This example sets its compute capability to $9.0$, running on an NVIDIA H100 GPU.
The \texttt{threadPerBlock} defines the GPU block dimensions,
and \texttt{template} specifies which code generation template to use. In this example, it is a global memory template.
The framework currently bundles the backends and their respective templates for sequential execution, OpenMP, CUDA, HIP, SYCL, and CSL programming models.
A later section describes each of them in detail.

To facilitate development, debugging, and evaluation, the \texttt{launch} construct takes additional optional parameters not shown in Listing~\ref{lst:stencilpy-star2d4r}.
For example, \verb|--print-code| outputs generated code to terminal,
\verb|--save-temps|  preserves all intermediate code in a subdirectory of \verb|/tmp|,
and \verb|--profile| measures the time spent in different framework components, including code parsing, generation, compilation, and execution.

\paragraph{Type Hints Required}

Python is a dynamically typed language.
While its types are only checked and validated at runtime, for us to generate correct code with our backends, type information needs to be known at compile time.
Furthermore, type information can be important in analyzing code for performance optimization opportunities.
Therefore, while type hints are optional in Python, StencilPy requires type hints in its \texttt{kernel} and \texttt{target} constructs.

\section{Framework Implementation and Optimization}
\label{sec:stencilpy-franework-impl}
The StencilPy framework includes components in   Python and C.
The frontend of the framework is implemented in Python,
handling the parsing of the DSL.
The rest of the framework, including optimizations and backend code generation, is implemented in C.
C language is known for its efficiency and simplicity, making it mostly suitable for efficient code execution while maintaining readability and ease of maintenance.
It operates at the right level for our purpose without introducing excessive processing overhead.

The interaction between Python and C code is facilitated using Python's $ctypes$ library and several dynamically-linked libraries compiled from the C code.
This approach enables the necessary functionality without introducing extra overhead that could impact execution performance.
Moreover, because $ctypes$ is built-in to Python, the framework doesn't require any extra third-party dependencies, facilitating the use of the framework on different platforms for better developer accessibility.

Unlike running kernels based on programming models such as OpenMP or CUDA, where the code needs to be compiled ahead of time (AOT), and then executed on the compiled binaries,
Python, being an interpreted language, enables users to run their code directly with Just-in-Time (JIT) compilation.
This shortens the feedback cycle and enables faster development iterations.
We want to preserve this user experience, so under the hood, despite the backend generating code that requires compilation, the framework employs JIT compilation.

While Python is portable, its interpreters are still platform-specific.
However, the details are transparent to end users.
Users only need to install the binaries for their platform and start using the tool.
StencilPy adopts the same approach to maximize portability and enhance user experience.
This also supports the common practice in scientific application development, where applications are initially developed on laptops or small-scale workstations, using a sequential backend or consumer-grade accelerators, and later evaluated on HPC clusters.

Allocating, accessing, and looping through large-size data grids in Python is not designed to be fast.
For example, in one of our development machines with a modern Intel Xeon CPU, populating a data grid of $1000^3$ with random 32-bit floating-point values in Python runtime takes approximately six minutes.
In response, StencilPy has its C-based data layer implementation for better performance.
Nevertheless, despite being implemented in C, the data layer is exposed as a Python module using $Python.h$ interface, seamlessly integrated into Python runtime with no third-party dependencies.

Furthermore, our data layer provides commonly used data array operations,
supplemented by stencil-specific operations.
These stencil-specific operations are handled differently depending on whether they operate on the host machine or the device.
When computing on a host, the framework directly executes the operators implemented in our data layer.
To compute on a device, the framework generates device-specific code.
Then, the framework compiles and executes the generated code on the target device.

The framework generates platform-specific code based on the provided backend configuration and, if needed, a C interface.
This code is then compiled and assembled into a shared library.
The shared library can then be loaded and executed with $ctypes$ by the framework.

\section{Stencil Performance Optimizations}\label{sec:stencilpy-opt}

To accelerate runtime performance, StencilPy's optimizers employ a range of parameters that control code analysis and generation.
While the long-term goal is to introduce an auto-tuner with a cost model to pick the best configurations, in the current version, users manually provide all these parameters, with certain parameters inferred by the framework.
For the inferred configurations, the framework allows them to be overwritten by users, providing flexibility in customization and can be helpful in cross-compilation scenarios.

\begin{table*}
\centering
\begin{tabular}{|p{1.5cm}|p{5.4cm}|p{6.6cm}|}
\hline
& \multicolumn{1}{c|}{Inferred by kernel definition}               & \multicolumn{1}{c|}{Available at the call site}                \\ \hline
Environ-ment parameters  &    & Device traits, such as device type, model, memory size, bandwidth, shared memory model or distributed memory model, memory hierarchy or flat memory, core frequency, etc); Host information; etc.  \\ \hline
Domain \newline parameters                  & Stencil orders; Stencil
 shapes, such as star, compact-in-space, and box; Stencil data array properties, such as number of elements, grid dimensions, memory locales, etc. & Problem domain size; PML layer width; Looping patterns; Number of iterations; etc.     \\ \hline
Perfor-mance \newline guidance \newline parameters & Domain decompositions, such as unified, two-region, and seven-region; Loop unrolling; Paddings to the innermost dimension and/or the 2D plane; global memory read with coalescing; etc                           &
\textit{CPUs}: OpenMP configurations, such as using loops, collapsed loops, tasks, looptask, etc. Applying Semi-stencil algorithm~\cite{de_la_cruz_introducing_2010,de_la_cruz_algorithm_2014}; \newline
\textit{GPUs}: Tiling strategies, such as 3D tiling and 2.5D tiling, and their respective parameters for variants, such as number of time steps, tile sizes; Algorithm choice, such as simple 3D mapping, streaming variants as in Nguyen~\cite{nguyen_35-d_2010}, Micikevicius~\cite{micikevicius_3d_2009}, and Matsuoka~\cite{matsumura_an5d_2020};
Semi-stencil~\cite{de_la_cruz_introducing_2010,de_la_cruz_algorithm_2014} on the streaming dimension; Data buffering related, such as whether to employ \verb|memcpy_async| if hardware supports it, number of planes to buffer; Hardware-specific features, such as vectorized data types (\verb|float4|); etc. \newline
\textit{STX}: Utilizing hardware features, such as plane-scheme. \newline
\textit{Cerebras}: PE private memory saving strategies; Using \textit{memcpy} for accelerating host-device data transfer; Using asynchronous communications for inter-PE data movements; vectorization of floating-point operations.
\\ \hline
\end{tabular}
    \caption{StencilPy performance-tuning parameters.}
    \label{table:stencilpy-perf-tuning-params}
\end{table*}

Table~\ref{table:stencilpy-perf-tuning-params} provides an overview of all the parameters supported by the framework.
These parameters can be categorized into three groups: environment parameters, domain-specific parameters, and user-guided performance-tuning parameters.
Most of these optimizations are detailed in previous chapters, and others will be discussed later when we discuss the intermediate representation design and code generations.

\section{Workflow}
\label{sec:stencilpy-workflow}

This subsection describes the framework's internal workflow -- how a piece of stencil code goes through our framework and eventually runs on the hardware.

The process begins with domain developers writing kernels in the DSL. The DSL code is parsed into an Abstract Syntax Tree~(AST). The AST is then transformed into the highest level of intermediate representations (IRs). While the AST is represented using Python's data structures, the intermediate representations use C-based data structures.

The IRs undergo an analysis phase to infer various properties of the stencil kernel. Static analyses are applied to understand the stencil computations and stencil data arrays.
In the first phases of IR transformations, the framework annotates the IR with additional information that represents the stencil kernel, such as stencil shape, looping pattern, stencil grid updates, local variables, and others, making it easier for optimization.
Next, the IR goes through multiple optimization phases.
Based on user-provided performance tuning parameters, the IR is manipulated and optimized using a suite of optimization techniques.
Then, architecture-specific information is utilized to optimize the kernel further. This process eventually generates an IR optimized for performance and designed to facilitate code generation.

Depending on the target architecture, the IR is fed into a platform-specific code generator. The code generator transforms it into target source code dedicated to a programming model or a particular architecture. For example:
\begin{itemize}
    \item For AMD, Apple, IBM, and Intel CPUs, \texttt{ompgen} generates C code with OpenMP directives;
    \item For Fujitsu A64FX, \texttt{ompgen} generates a specialized OpenMP code with configurations dedicated to the Fujitsu compiler;
    \item For NVIDIA GPUs, \texttt{cudagen} generates CUDA code and its C interface;
    \item For AMD GPUs, \texttt{hipgen} generates HIP code and C interface;
    \item For Intel GPUs, \texttt{syclgen} generates SYCL code;
    \item For STX, \texttt{stxgen} generates OpenMP code with STX's custom OpenMP extension;
    \item For Cerebras, \texttt{cslgen} generates CSL code and its Python runner.
\end{itemize}
In addition, the framework also has a basic code generator \texttt{cgen} to generate serial code in C.

Finally, the backend-specific runner compiles and executes the generated code.
For all but Cerebras backends, the runners invoke platform-dedicated compilers to build and assemble the code into shared libraries.
The shared libraries are then loaded and executed.
For Cerebras, the framework generates multiple CSL files for stencil computations, task and state management, and fabric layout. These CSL files are compiled into multiple executable files in ELF using \texttt{cslc} compiler. Finally, \texttt{cs\_python}, a Cerebras-specialized Python, distributes these ELFs onto either Cerebras hardware or simulators and moves data between host and device for execution.

After the execution, the results are stored in the memory spaces in our data layer so that the program can inspect them or provide them to enclosing code.

\section{Backend Templates}
\label{sec:stencilpy-backend-templates}

StencilPy supports seven backends, namely \texttt{seq} (for sequential code), \texttt{omp} (for OpenMP), \texttt{cuda}, \texttt{hip}, \texttt{sycl}, \texttt{stx} (for STX-specific OpenMP code generation), and \texttt{csl} (for Cerebras CS-2).
The framework includes template(s) specific to each backend, serving as the basis for code version variations.
Depending on the domain problem and backend, developers can experiment with templates and their variants for the best performance.
Most templates come with parameters, allowing for customization.
Developers can tweak these ``knobs'' to tune the performance.

\subsection{Sequential Backend (seq)}

The most basic backend executes the DSL in a single-threaded fashion using a single CPU core. It supports most CPU architectures, such as ARM, IBM POWER, and x86. It provides one template with no parameters. The main purpose of offering such a template in the framework is to establish a baseline for numeric results and sequential execution performance.

\subsection{OpenMP Backend (omp)}

The framework provides OpenMP backend for code versions utilizing OpenMP programming models.
It offers templates for two stencil shapes: conventional stencil computation and Semi-stencil, respectively.
For each of the stencil shapes, the backend provides five OpenMP configurations as depicted in Table~\ref{table:stencilpy-omp-templates}.
In total, StencilPy can produce OpenMP code variants of ten combinations of two stencil shapes and five OpenMP configuration strategies.

\begin{table*}[t]
\centering
\begin{tabular}{|c|m{12cm}|}
\hline
Configuration        & \multicolumn{1}{>{\centering\arraybackslash}c|}{Description}                                             \\ \hline
\texttt{loop}           & Uses a \texttt{runtime} schedule and applies an OpenMP \texttt{parallel for} to the outermost loop.  \\ \hline
\texttt{loop\_blocking}       & Adopts a \texttt{runtime} schedule and uses blocking in the outermost two dimensions with
OpenMP \texttt{parallel for} applied to the 2D loop nest over the blocks.      \\ \hline
\texttt{loop\_blocking\_collapse}       & In addition to the configuration of loop\_blocking, applies \texttt{collapse(2)} to combine the two loops.  \\ \hline
\texttt{tasks\_blocking}       & Uses blocking in the outermost two dimensions and runs each block as an OpenMP \texttt{task}. \\ \hline
\texttt{taskloop}       & Applies a \texttt{taskloop} to the loops, runs each chunk of iterations as a \texttt{task}. \\ \hline

\end{tabular}
    \caption{Strategy templates for OpenMP backend on a 3D stencil.}
    \label{table:stencilpy-omp-templates}
\end{table*}

\sloppy
Table~\ref{table:stencilpy-omp-templates} outlines the OpenMP configurations, and for the ones using blocking strategies (\texttt{loop\_blocking}, \texttt{loop\_blocking\_collapse}, and \texttt{tasks\_blocking}), StencilPy provides parameters to configure the sizes of the 2D block dimensions.

These code versions and their optimizations are intended to support modern many-core CPUs, such as the HPC nodes sporting AMD EPYC GenoaX, Fujitsu A64FX, IBM Power9, and Intel Sapphire Rapids. Besides, they can be used to speed up the local development of kernels on multi-core CPUs, such as AMD Ryzen, Apple Silicon, and Intel Core.

To cater to a diverse range of platforms, the OpenMP backend carries optimal compilation configurations, including the type of compiler, compiler frontend flags, and linker flags. The default values of these configurations are provided by the framework's frontend, ensuring that the backend utilizes them to achieve the best performance on the specific platform. However, users can overwrite these configurations or provide new configurations as part of the backend clause.

\subsection{GPU Backend (cuda, hip, and sycl)}

GPU vendors have their respective programming models, but they are similar for the most part, although the details differ.
Table~\ref{table:gpu-templates} presents all the built-in templates for 3D stencils tailored on GPU backends, including CUDA, HIP, and SYCL.

\begin{table*}[t]
\centering
\begin{tabular}{|c|p{12.6cm}|c|}
\hline
Template        & \multicolumn{1}{c|}{Description}                                             & Blocking                         \\ \hline
gmem           & 3D blocking using global memory only                    & \multirow{3}{*}{3D}     \\ \cline{1-2}
smem       & 3D blocking using shared memory for stencil grids        &                                                      \\ \cline{1-2}
f4      & 3D blocking using global memory and \verb|float4| data type for vectorized computations    &                                                       \\ \cline{1-3}
shift       & 2.5D streaming, data on the streaming dimension are ``shifted'' when streaming to the next plane    & \multirow{3}{*}{2.5D}                                 \\ \cline{1-2}
unroll  & 2.5D streaming, data on the streaming dimension are ``fixed'' using loop unrolling                   &                                 \\ \cline{1-2}
semi & 2.5D streaming using Semi-stencil algorithm &                        \\ \hline
\end{tabular}
    \caption{Built-in templates for GPU backends on a 3D stencil.}
    \label{table:gpu-templates}
\end{table*}

3D-blocking templates parameterize block size along the $X$, $Y$, and $Z$ axes.
Similarly, templates of 2.5D blocking parameterize plane sizes for their $X$ and $Y$ dimensions.

When specified using \verb|--mem-type| flag, 2.5D-blocking templates generate code using registers or shared memory for stencil points on the streaming dimension.
Furthermore, when no memory type is specified, the framework automatically selects the memory type based on the stencil shape,
with star-shaped stencils employing registers and other stencil shapes employing shared memory.
This is possible because the framework identifies the stencil shape in the analysis and optimization phase.

In addition, prefetching can be enabled for 2.5D templates using \texttt{--prefetch} flag, enabling data fetching to overlap with computations.
Lastly, when hardware support is present, \texttt{--asyncmemcpy} flag can be set to move data without staging through registers, freeing registers to be used by computations. This option is currently available to CUDA backend that runs on the latest generations of NVIDIA GPUs.

The GPU templates also support 2D stencils with everything described but one dimension lower.

\subsection{STX Backend (stx)}

The framework comes with three templates for STX: \verb|cube|, \verb|plane|, and \verb|semi|.
STX templates are limited to 3D stencils due to STX toolchain restrictions.
The \verb|cube| template resembles GPU's 3D blocking.
The \verb|plane| template aligns with STX's plane scheme.
The \verb|semi| template applies the Semi-stencil algorithm on top of the plane scheme.
Both cube sizes and plane sizes can be adjusted with parameters for performance tuning.

\subsection{CSL Backend (csl)}

There are unique features in the dataflow architecture of the Cerebras CS-2 compared to conventional platforms. As a result, hand-crafting CSL code is time-consuming and error-prone, and many optimizations employ CSL's special features that harm code readability and make it harder to maintain. To our knowledge, StencilPy's CSL backend is the first attempt to address these issues, enabling developers to write high-level stencil code and use a framework to handle the CSL code generation and optimization for the CS-2. Although CSL backend has a single template with no parameter, it significantly lowers the barrier to using a Cerebras system for stencil-based computations.

\bigskip

\noindent
We believe the framework's built-in templates for each backend cover a comprehensive range of high-performance algorithmic approaches and optimizations for high-order stencils, drawing inspiration from prior research.
Nonetheless, should the built-in templates not align with the needs of a specific domain application,
developers can extend the framework by introducing new templates to the backend, thanks to the backend template system's modular design.
When working with a new architecture, it is also quite straightforward to introduce templates for backends and ``plug-in'' them into the framework.
Throughout StencilPy's development, we have applied this methodology to developing all backends.

\section{Intermediate Representations}\label{sec:stencilpy-ir}

Many of StencilPy's features, such as code generation support for a wide range of backends using various templates and performance optimizations, rely on static analysis. StencilPy's static analysis comprehends the code and considers it from multiple perspectives.
To facilitate static analysis in StencilPy, it employs a multi-layer intermediate representation (IR).

Unlike  IR designs used by compiler toolchains for general-purpose languages that must preserve source locations and maintain higher-level semantics, the sole goal of StencilPy's IR is to support the generation of highly performant code that runs on modern node architectures.

\begin{figure*}
    \centering
    \includegraphics[width=0.8\textwidth, trim={0 4.5cm 0 3cm}, clip]{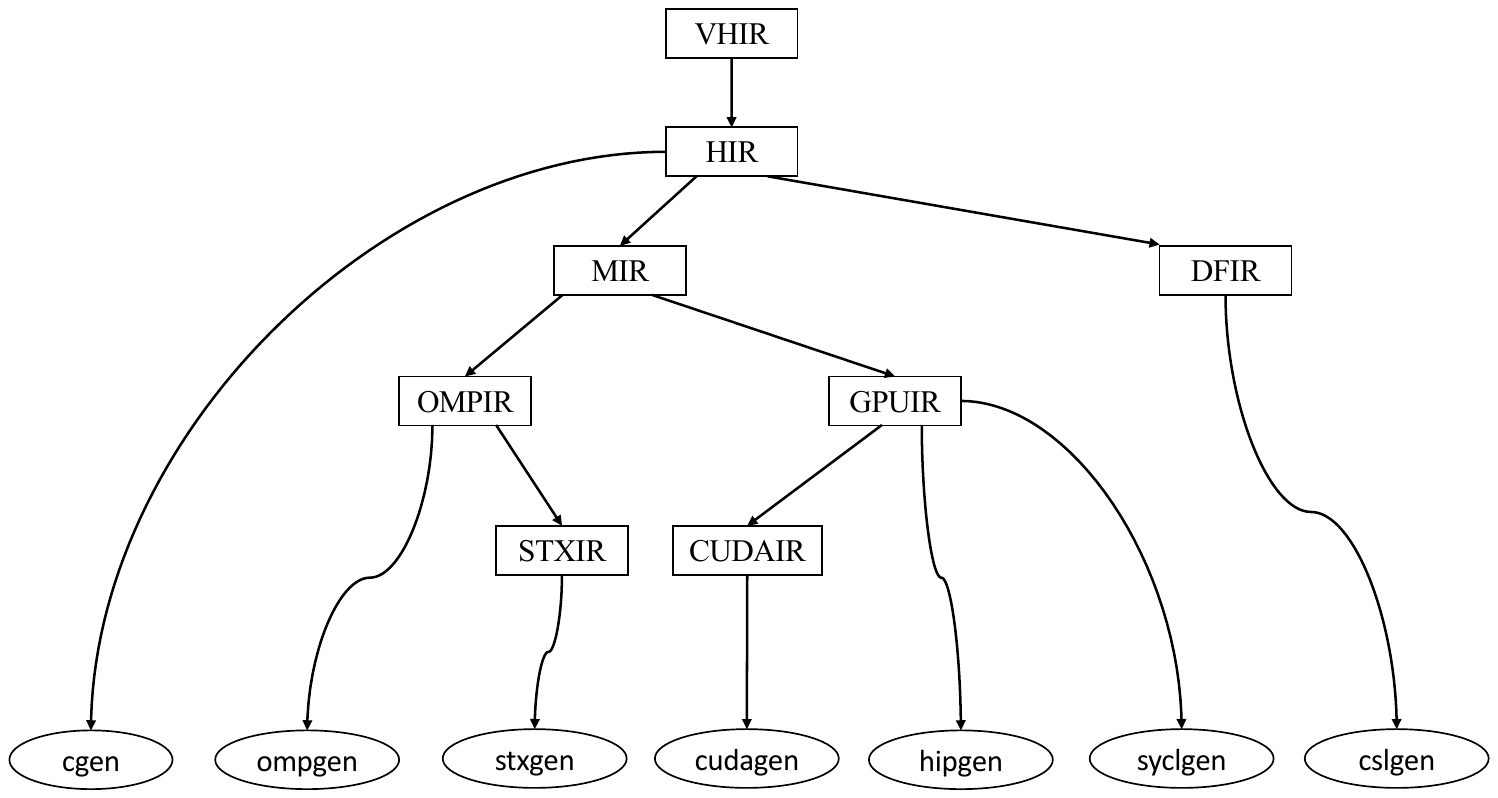}
    \caption{IR hierarchy in StencilPy framework.}
    \label{figure:stencilpy-ir-hierarchy}
\end{figure*}

Figure~\ref{figure:stencilpy-ir-hierarchy} depicts our IR hierarchy.
Higher-level IRs are closer to the syntactic abstraction of the DSL, while lower-level IRs are closer to machine code abstractions.
In each static analysis phase, the framework transforms these IRs from higher to lower.

\subsection{VHIR}

VHIR sits at the top of the hierarchy and represents the same information as the DSL's AST. This is the interface between the framework's Python models and C data structures.

\subsection{HIR}

The framework traverses the VHIR, extracts the information about the stencil involved in computation and its data domain, and yields an HIR.

At this level, the framework analyzes data regions, identifying the boundaries of data subregions.
The framework examines stencil computations to understand their radius and data grid dimensions.
Synthesized from grid data and PML sizes, the framework infers the loop details, including the number and size of loop dimensions and the looping patterns.
Furthermore, it comprehends the stencil patterns related to point locations by identifying index offsets involved in computations, assessing the dimensionality of the stencil, and determining whether those points are perfectly aligned on the cardinal axes or not.

HIR is used directly by \texttt{cgen}, which generates serial code.

\subsection{MIR}

MIR incorporates annotations for blocking strategies, such as whether 3D stencils should employ 3D blocking or 2.5D blocking, and for 2D stencils, whether to utilize 2D blocking or 1.5D blocking. MIR maintains a lightweight symbol table, recording local variables and the information around the variables, such as its position at the kernel's top level, whether its right-hand side involves a stencil update, whether it functions as a temporary variable, or if it serves as the destination for a stencil update.

While no code generator directly consumes MIR, it serves as a base IR for both GPUIR and OMPIR and contains shared information used by both children IRs.

\subsection{GPUIR}

GPUIR concerns details related to GPU-specific hardware features, such as global memory, shared memory, constant memory, and registers. It determines optimal strategies for data storage, access, and movement among these memory types on a GPU device.

In cases where the prefetch configuration is on, GPUIR calculates the additional memory buffer necessary for prefetching.

Additionally, when the specified GPU sports vectorization units, GPUIR transforms regular floating-point types into \texttt{float4} data types for efficient vectorization.

GPUIR possesses sufficient information suitable for direct input into \texttt{hipgen} and \texttt{syclgen}, while \texttt{cudagen} can use some extra information for its processing.

\subsection{CUDAIR}

NVIDIA has introduced an asynchronous memory copy feature in their latest A100 and H100 GPUs, enabling data movement without staging through registers. Since asynchronous memory copy doesn't need registers during data prefetching, they can be utilized for overlapping computations concurrently, leading to improved performance. When asynchronous memory copy is enabled, CUDAIR notes it to take advantage of these features.

To accommodate all recent generations of NVIDIA GPUs, GPUIR retains information about the current device's compute capability. This ensures that \texttt{cudagen} employs the new asynchronous memory copy feature when supported by the device and falls back to the conventional approach when it is not.

\subsection{OMPIR}

OMPIR understands looping patterns, including the number of dimensions and whether it directly loops on data grid dimensions. Additionally, it considers whether and how to iterate over a block when accounting for variations in blocking strategies when \texttt{loop\_blocking}, \texttt{loop\_blocking\_collapse}, and \texttt{tasks\_blocking} templates are employed.

OMPIR is the input for \texttt{ompgen}. It also serves as a basis for STXIR.

\subsection{STXIR}

The STX programming model extends OpenMP with architecture-specific extensions. OpenMP still manages computations on its host CPU, while the extension introduces nested offloading, specifically addressing offloading tasks to STX processing units.~\cite{ryuichi_rice_energyhpc_2023} STXIR annotates the on-device computation and inserts device memory allocation and data movement between the host and device. With STX-specific information annotated, STXIR is ready to be consumed by \texttt{stxgen}.

\subsection{DFIR}

Dataflow architecture has unique features, so DFIR needs to handle many aspects, such as data mapping on a processing element (PE), data movement between host memory and device memory, data communication among PEs, and vectorization of the stencil computation. To coordinate computation and distributed asynchronous data communications, synthesized from the stencil computation, DFIR dynamically constructs a state machine for each program execution.
Each state machine defines information such as the data preparation, communication, and computation for each stencil point, the current state, and the next state for each state. The state machine receives asynchronous callbacks and decides when to move to the next state. DFIR also introduces stencil point index patterns to facilitate the management of the state machine and data dependency. We delve into stencil point index patterns in detail in this subsection, while deferring the discussion on state machine generation to when we describe code generation for the CSL backend.

Derived from HIR with additional stencil and data grid-related information, DFIR is annotated with details such as the symbol list and stencil point index patterns. The symbol list is crucial for exposing the symbol interface, enabling data flow between the host and the device. Stencil point index patterns provide essential information about the offset, distance, and location relationships in stencil computations, influencing state management, data communication patterns, and vectorization.

We annotate DFIR with hardware information, including fabric dimensions, requested PE width and height based on stencil grid size, and the layout of activated PEs within the fabric. We also allocate buffer PEs for data to flow from and to the host.

Typically, it is straightforward to incorporate the total time iterations into the target. However, in the context of dataflow architecture, where total time iterations and current iteration are integral parts of state machine management, it is required for the total time steps to be known during compilation, thus, is recorded in DFIR.

\begin{figure}
  \centering
  \includegraphics[width=0.42\linewidth]{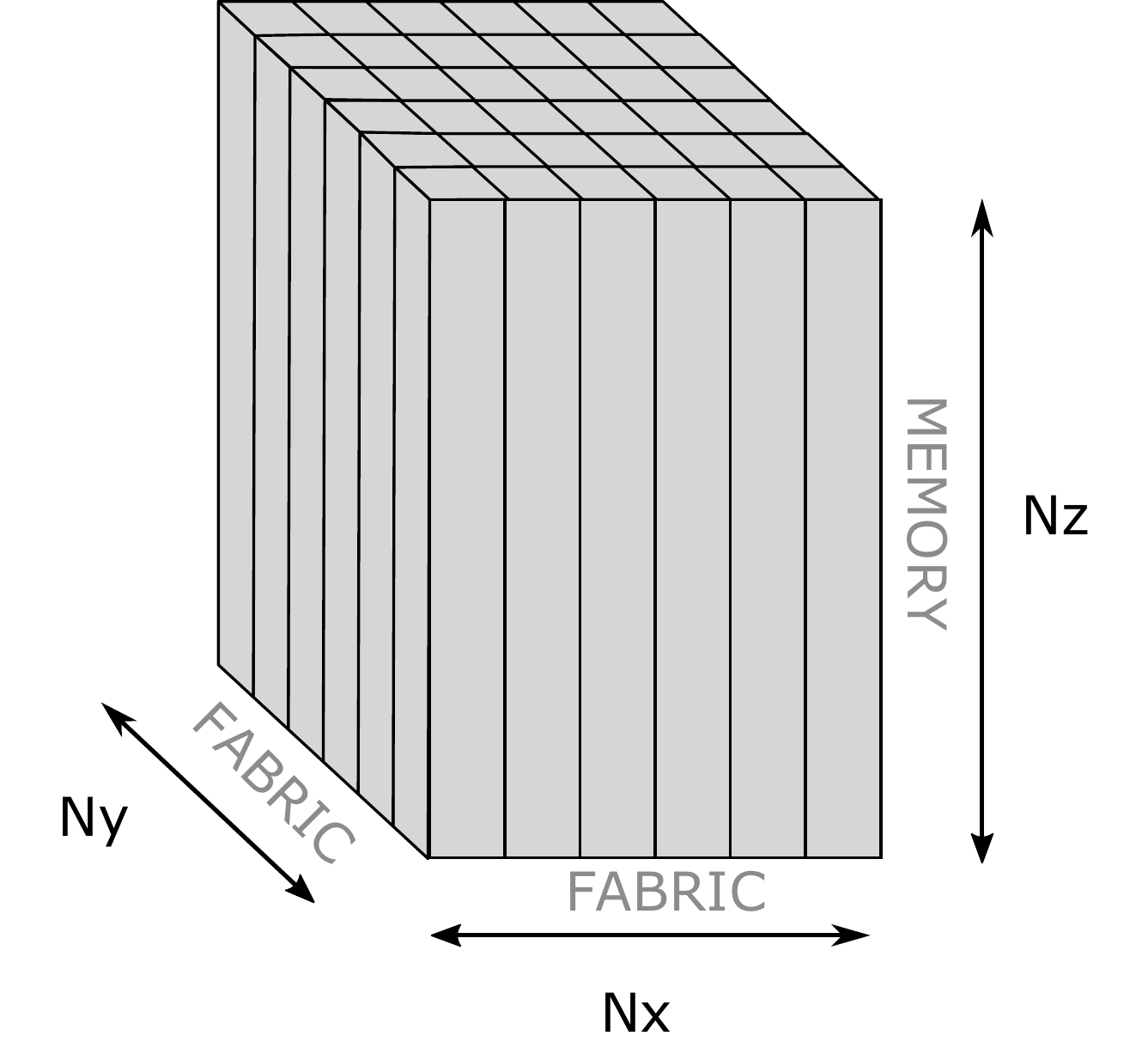}
  \caption{3D grid of size ${N_x \times N_y \times N_z}$. X and Y dimensions are mapped onto the PE grid of the WSE, while Z dimension is mapped onto memory of each PE.}\label{figure:pe-mapping}
\end{figure}

Stencil grid mapping onto a WSE wafer involves decomposing the data domain, with each Z-dimension cell mapped to the same PE, and X and Y dimensions spread across the wafer axes as shown in Figure~\ref{figure:pe-mapping}.
The whole Z-dimension resides in the local memory of a PE.
This strategy, as demonstrated in previous work~\cite{DirkSC20,MauricioSC22,ryuichi_scalah_2023}, optimizes the utilization of dataflow architecture and maximizes the potential parallelism for HPC applications of this kind.
The stencil grid's width and height are constrained by maximum PE dimensions, while its depth is limited by PE's private memory size. So, they are also annotated in DFIR.

\begin{figure}[t]
    \centering
    \includegraphics[width=0.6\textwidth, trim={3.3cm 3cm 3.6cm 3cm}, clip]{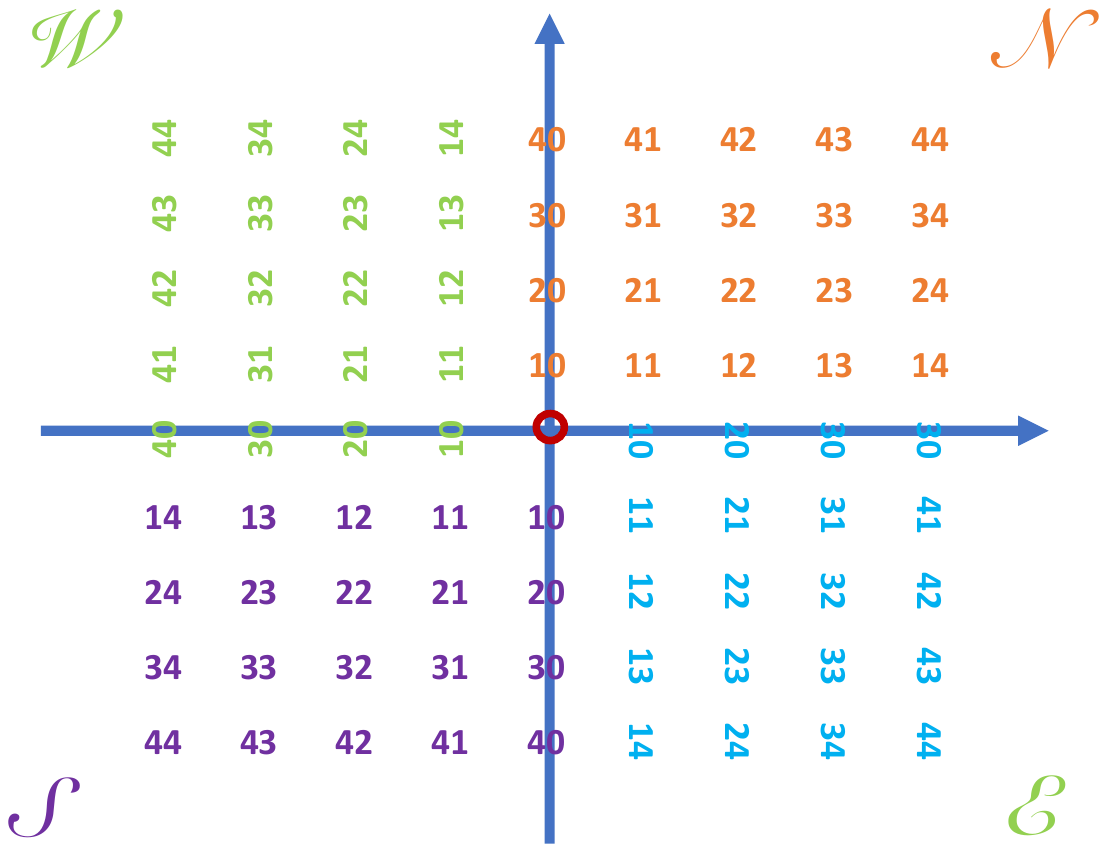}
    \caption{Stencil point index pattern used in DFIR and CSL code generation.}
    \label{figure:stencilpy-csl-stencil-point-index-pattern}
\end{figure}

In Figure~\ref{figure:stencilpy-csl-stencil-point-index-pattern}, we present the structure of our stencil point index patterns for a stencil with a radius of four. The center position is labeled as pattern identifier \texttt{0}. Other positions in each quadrant are uniquely identified based on their offset from the center point. By analyzing the stencil pattern in computations, we map the stencil offset index to the corresponding stencil point index pattern identifier. Each quadrant is labeled with a directional tag. For instance, quadrant I is denoted as $N$, II as $W$, III as $S$, and IV as $E$. The identifiers for each pattern do not carry numeric significance. Instead, they are labeled to enhance readability --- the tens digit indicates the offset along the axis from the center point, while the ones digit indicates the distance from the perpendicular axis. Rotating the index patterns counterclockwise in the $N$ quadrant provides the index patterns for the other three quadrants. Moreover, the indices associated with each axis are linked to a specific quadrant. Algorithm~\ref{algo:stencilpy-offset-index-to-pattern-id} outlines the procedure for mapping a stencil point's offset to a stencil point index pattern identifier. Parallel data communication can be achieved for the stencil points sharing the same index pattern identifiers from different quadrants.

\begin{algorithm}[t]
    \KwIn{\\\Indp \Indp
    $\mathbf{x, y, z}$: index offset from center point of x, y, and z dimension, respectively}
    \KwOut{\\\Indp \Indp
    $\mathbf{id}$: stencil point index pattern identifier}

    \eIf{$\mathbf{x} = 0 \land \mathbf{y} = 0$}{
        $ \mathbf{id} \gets 0$ \;
    }{
        \Switch{$(\mathbf{x}, \mathbf{y})$} {
            \lCase{$( >= 0, > 0 )$}{$ d \gets N, i \gets \mathbf{y}, j \gets \mathbf{x} $}
            \lCase{$( > 0, <= 0 )$}{$ d \gets E, i \gets \mathbf{x}, j \gets -\mathbf{y} $}
            \lCase{$( <= 0, < 0 )$}{$ d \gets S, i \gets -\mathbf{y}, j \gets -\mathbf{x} $}
            \lCase{$( < 0, >= 0 )$}{$ d \gets W, i \gets -\mathbf{x}, j \gets \mathbf{y} $}
        }

        $\mathbf{id} \gets d \oplus i \oplus j $ \tcc*[r]{$\oplus$ denotes string concatenation}
    }
    \caption{Mapping stencil offset index to a stencil point index pattern identifier.}
    \label{algo:stencilpy-offset-index-to-pattern-id}
\end{algorithm}

\begin{figure}[t]
    \centering
    \includegraphics[width=0.3\textwidth, trim={10.5cm 7.8cm 10.5cm 7cm}, clip]{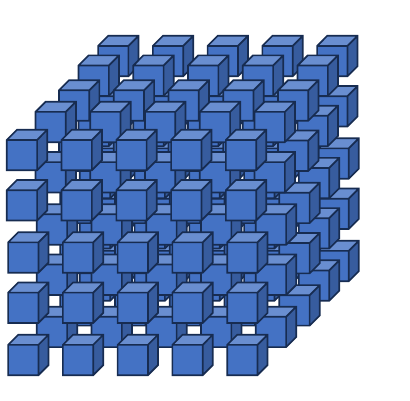}
    \caption{Shape of a 3D box-shaped stencil with a radius of two.}
    \label{figure:stencilpy-box3d2r-example}
\end{figure}

\begin{figure}[t]
    \centering
    \includegraphics[width=0.5\textwidth, trim={6cm 7cm 6cm 7cm}, clip]{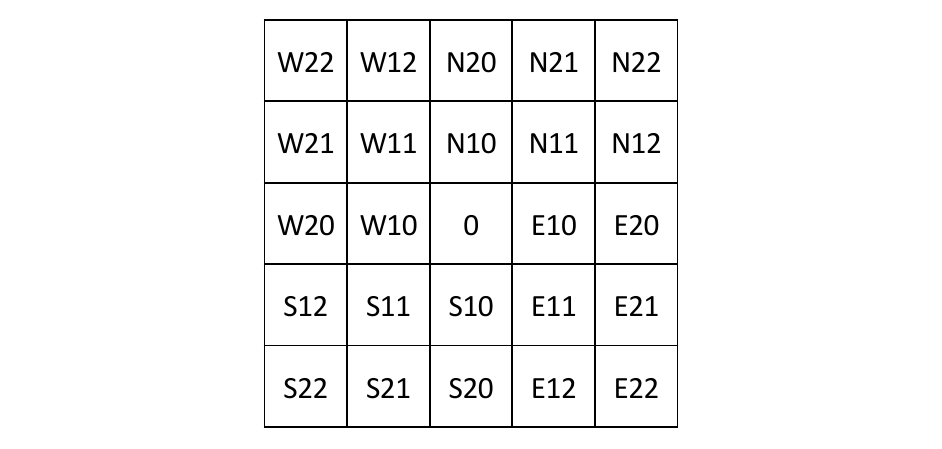}
    \caption{Stencil point index pattern identifiers for the stencil in Figure~\ref{figure:stencilpy-box3d2r-example}.}
    \label{figure:stencilpy-box3d2r-index-patterns}
\end{figure}

Subsequently, we provide an example to illustrate the concept of stencil point index pattern and its mapping algorithm. Figure~\ref{figure:stencilpy-box3d2r-example} shows the stencil points in a 3D box-shaped stencil with a radius of two. Upon applying Algorithm~\ref{algo:stencilpy-offset-index-to-pattern-id}, the resulting mapping is depicted in Figure~\ref{figure:stencilpy-box3d2r-index-patterns}.

The stencil point index pattern is formulated in 2D to conceptually align with the fabric's 2D layout. As the entire Z-dimension is contained within the local memory of a PE, the stencil point index pattern is annotated with the maximum Z-dimension offset. Algorithm~\ref{algo:stencilpy-pattern-id-z-max-offset} describes the process of annotating a stencil point index pattern with its associated maximum Z-dimension offset.

\begin{algorithm}[t]
    \KwIn{\\\Indp \Indp
        $\mathbf{a}$: an array of stencil offset indexes\\
        $\mathbf{id}$: stencil point index pattern identifier
    }
    \KwOut{\\\Indp \Indp
        $\mathbf{m}$: maximum Z-dimension offset for the provided $\mathbf{id}$
    }

    $\mathbf{m} \gets 0$ \;

    \ForEach{$(x,y,z)$ in $\mathbf{a}$}{
        $pid \gets $ apply Algorithm~\ref{algo:stencilpy-offset-index-to-pattern-id} with $(x,y)$ \;
        \If{$pid = id$} {
            \If{$z > m$} {
                $\mathbf{m} \gets z$ \;
            }
        }
    }
    \caption{Annotating a stencil point index pattern identifier with its associated maximum Z-dimension offset.}
    \label{algo:stencilpy-pattern-id-z-max-offset}
\end{algorithm}

\begin{algorithm}[t]
    \KwIn{\\\Indp \Indp
    $\mathbf{a}$: an array of stencil offset indexes}
    \KwOut{\\\Indp \Indp
    $\mathbf{s}$: sorted dependency array of $\mathbf{a}$}
    \SetKwFunction{FMain}{DependsOn}
    \SetKwProg{Fn}{Function}{:}{}
    \Fn{\FMain{$id1, id2$}}{
        $ dep \gets false $ \tcc*[r]{if $id1$ depends on $id2$, return true}

        $d1 \cdot i1 \cdot j1 \gets id1 $ \;
        $d2 \cdot i2 \cdot j2 \gets id2 $ \tcc*[r]{extracts d, i, j from id}

        \eIf{$ d1 \neq d2 $}{
            $ dep \gets false$ \tcc*[r]{patterns from different quadrants never depend on each other}
        }{
            \eIf{$ j1 = j2 $}{
                $ dep \gets i1 > i2$ \;
            }{
                $dep \gets j1 > j2$ \;
            }
        }
        \textbf{return} $ dep $ \;
    }
    \textbf{end}

    $\mathbf{s} \gets $ sort \textbf{a} using \verb|DependsOn| as the comparison function \;

    \caption{Constructing dependencies among stencil point index patterns.}
    \label{algo:stencilpy-pattern-id-dependency}
\end{algorithm}

\begin{figure}[t]
    \centering
    \includegraphics[width=0.45\textwidth, trim={7.2cm 4cm 7.2cm 3cm}, clip]{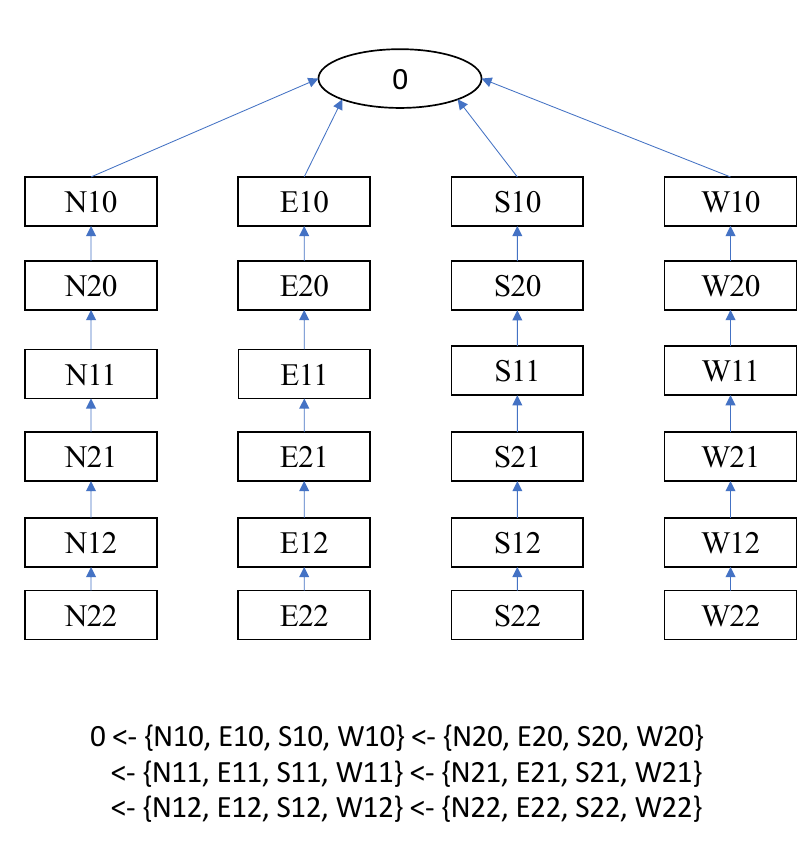}
    \caption{Stencil point index pattern dependency for the stencil in Figure~\ref{figure:stencilpy-box3d2r-example}.}
    \label{figure:stencilpy-box3d2r-deps}
\end{figure}

Because a PE router only connects to the routers of its four cardinal directly neighboring PEs, for data communication, pattern \texttt{10} in any quadrant can send data to the center point using the router links directly. In contrast, all other patterns need to indirectly send data to the center point, relying on one or more other patterns from the same quadrant to complete prior to its own communication. For example, pattern \texttt{N20} depends on pattern \texttt{N10}, pattern \texttt{E30} depends on pattern \texttt{E20} then indirectly depends on pattern \texttt{E10}, and pattern \texttt{W22} cascadely depends on patterns \texttt{W21}, \texttt{W20}, and \texttt{W10}. Consequently, depending on the data's location in the stencil computation, DFIR conducts data analysis, constructing a dependency array using the steps described in Algorithm~\ref{algo:stencilpy-pattern-id-dependency} for data communications from these stencil point index patterns. Figure~\ref{figure:stencilpy-box3d2r-deps} provides the dependency graph for the example stencil depicted in Figure~\ref{figure:stencilpy-box3d2r-example}. This dependency array is subsequently used to generate code that controls the router to choreograph the sending and receiving of data. It also manages the state machine and tracks progress to ensure that data communications adhere to the specified dependency order.

Finally, DFIR transforms the numeric operations in a stencil computation from a tree data structure into a Static Single Assignment (SSA)~\cite{compiler_ssa} form based on stencil point index patterns. It is needed for generating vectorized code.

DFIR is the input to \texttt{cslgen}, which generates CSL code.

\section{Code Generation}\label{sec:stencilpy-codegen}

Our code generators produce platform-specific source code, leveraging platform compilers to generate machine code tailored to the target architecture. For instance, \texttt{cudagen} generates CUDA code, using \texttt{nvcc} to produce the corresponding NVIDIA GPU machine code. Similarly, \texttt{cslgen} generates CSL code, and \texttt{cslc} is employed to generate an ELF executable for Cerebras systems.
Given our framework's goal of transforming a user-friendly stencil DSL into efficient code, it is best to reuse existing platform-specific infrastructure instead of reinventing the wheel.
Therefore, our code generators produce the code in such a way that allows us to delegate certain traditional compiler optimizations to the platform-specific compilers, enabling us to concentrate on stencil optimizations.

In general, each code generator accepts an IR as input and produces code for stencil kernels as well as target code for launching the kernels. They generate code to map the data region to the kernel where the stencil computation occurs. They also handle memory allocation on the host.
Additionally, some code generators generate C interfaces for interaction with the framework frontend.

\subsection{Serial Code (cgen) and Commonly Shared Code Generation}

\texttt{cgen} takes HIR and generates serial versions of the code in C. While this process is straightforward, it serves as an opportunity to extract commonly shared code generation logic that can be reused in other code generators.

For instance, the generation of primitive numeric types, such as \texttt{int} and \texttt{float}, can be reused across different code generators. Similarly, the generation of most numeric operators (including prefix and binary operators) and control flow logics (such as \verb|if-else| statements and \verb|for| statements) can be shared among all code generators except \texttt{cslgen}.

Regarding \verb|for| statements, the DSL adopts Python's for-range loop syntax, so the framework converts it to a C \verb|for| loop with an initial condition, termination condition, and incremental statement.

Shared logic also includes stencil index arithmetic. The DSL defines stencil points based on offsets from an implicitly defined current stencil point index, while its data structure is flattened to a 1D data array in host memory. Halo regions are part of the physical memory layout, so the arithmetic between the two indexing systems is implemented here and can be reused.

\begin{figure}
    \centering
    \includegraphics[width=0.39\textwidth,trim={6cm 4cm 6cm 3.6cm},clip]{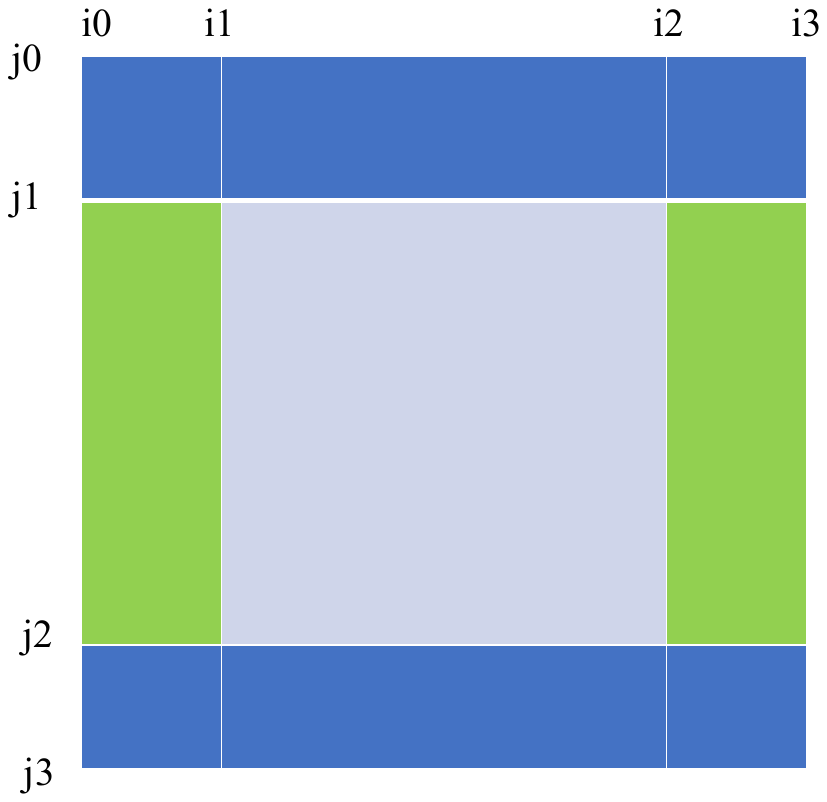}
    \caption{2D decomposition defined with \texttt{map} construct.}
    \label{figure:stencilpy-map-2d-decomp}
\end{figure}

\sloppy
Additionally, the logic for handling data decomposition and launching kernels according to decomposed data regions can be reused. Figure~\ref{figure:stencilpy-map-2d-decomp} illustrates a 2D decomposition using boundary indices as defined in DSL code \verb|map(i=(i0,i1,i2,i3),j=(j0,j1,j2,j3)|. The framework provides syntactic sugar to accommodate common use cases, such as
\begin{itemize}
    \item \verb|map(i=x,j=y)| is translated to \verb|map(i=(0,0,x,x),j=(0,0,y,y))|;
    \item \verb|map(i=x,j=y,w=p)| to \verb|map(i=(0,p,x-p,x),j=(0,p,y-p,y))|;
    \item \verb|map(i=(x1,x2),j=(y1,y2))| to \verb|map(i=(x1,x1,x2,x2),j=(y1,y1,y2,y2))|;
    \item \verb|map(i=(x1,x2),j=(y1,y2),e=p)| to \verb|map(i=(x1,x1+p,x2-p,x2),j=(y1,y1+p,y2-p,y2))|;
    \item \verb|map(e=(x,y))| to \verb|map(i=(0,0,x,x),j=(0,0,y,y))|; and
    \item \verb|map(e=(x,y),w=p)| to \verb|map(i=(0,p,x-p,x),j=(0,p,y-p,y))|.
\end{itemize}
Although parentheses are used in the DSL syntax, it's important to note that all intervals are semantically left-closed and right-open. The 3D decomposition is similar but includes a third parameter \texttt{k} to define the decomposition for the Z-dimension.

\subsection{OpenMP Code Generation (ompgen)}

In addition to generating common workflow and reused code, \texttt{ompgen} primarily focuses on three aspects of OpenMP-specific code generation: kernel code, blocking strategies, and various OpenMP pragmas for different configurations.

For kernel code, \texttt{ompgen} generates one of two code versions based on the backend parameter: one using conventional stencil computation and the other applying the Semi-stencil algorithm.

Depending upon the blocking strategies implied by the chosen template, \texttt{ompgen} either produces code with looping patterns that iterates through the entire data grid or decomposes the data grid into blocks and loops over the block dimensions. The loop dimension lengths are determined by the data grid, while the block dimensions are provided as backend template parameters.

The selected template also determines how \texttt{ompgen} incorporates OpenMP pragmas into the code.
\begin{itemize}
    \item \verb|#pragma omp parallel for| \verb|default(shared) schedule(runtime)| is applied to all loop-based templates, including \texttt{loop}, \texttt{loop\_blocking}, and \texttt{loop\_blocking\_collapse}. The \texttt{loop} template applies them to the outer loop, while \texttt{loop\_blocking} and \texttt{loop\_blocking\_collapse} apply them to 2D blocks. \texttt{loop\_blocking\_collapse} further applies \verb|collapse(2)| to collapse 2D blocks;
    \item For the \texttt{tasks\_blocking} template, \verb|#pragma omp parallel default(shared)| and \verb|#pragma omp master| are applied to the outer loop over time iterations, where \verb|#pragma omp task| is applied to the inner loop over 2D blocks. \texttt{ompgen} also adds a \verb|#pragma omp taskwait| at the end of the timestep; and
    \item if the \texttt{taskloop} template is used, \verb|#pragma omp parallel default(shared)|, \verb|#pragma omp single|, and \verb|#pragma omp taskloop| are applied to the outer loop.
\end{itemize}

\subsection{GPU Code Generation (cudagen, hipgen, and syclgen)}

While GPUs from the three major vendors have their respective programming models, GPU code generation shares some commonalities. We describe the common elements applicable to all three GPU code generators, and when variations arise, we address their differences.

The initial step in GPU code generation involves generating code for the allocation of data arrays in on-device memory. The stencil data grid is flattened into a 1D array layout. Padding between rows and lead padding to adjust the initial offset of each row are added to the memory layout to align inner region blocks with cache lines. Subsequently, code is generated to copy data from the host to the device. The generated data facilitates data copying with memory coalescing.

Next, code generators create GPU kernels and generate the corresponding kernel launch calls. In addition to launching kernels with the data decomposition previously described, when offloading kernels to GPUs, they generate code to instruct the allocations of GPU threads and threadblocks, as part of the kernel launch calls. GPU threadblocks are determined by the blocking strategies implied by the chosen template.

Both \texttt{cudagen} and \texttt{hipgen} generate code to allocate \texttt{streams} for CUDA and HIP, respectively, and launch kernels on these \texttt{streams}. On the other hand, \texttt{syclgen} produces a \texttt{submit} call and uses its command group handler to dispatch kernels onto devices.

The generation of kernel code is primarily decided by the kernel definition, selected template, and its parameters. In templates using 3D blockings, the first step involves determining the stencil center point index that the current thread handles. For both CUDA and HIP, this computation relies on implicitly-defined variables such as \texttt{blockIdx}, \texttt{blockDim}, and \texttt{threadIdx}, while \texttt{syclgen} can easily obtain the index from a \texttt{nd\_item}.

For the \texttt{gmem} template, code generation is straightforward. The generated code retrieves stencil point values directly from device memory, performs the computation, and directly updates the results in the device memory.

In the case of the \texttt{smem} template, the process requires several steps. First, code is generated to allocate shared memory space for CUDA and HIP, and \texttt{local\_accessor} for SYCL. Subsequently, stencil indices in both global memory layout and shared memory layout are generated. Following this, code is generated to load data from global memory to shared memory. Finally, stencil computation code is generated with data being read from shared memory. The updated stencil value is then updated in device memory.

The \texttt{f4} template is similar to \texttt{gmem} in that it retrieves data directly from global memory. However, the key difference lies in each GPU thread concurrently computing four floating-point numbers using its vector engine. The code generation involves converting the stencil index to the components of \texttt{float4} (\texttt{x}, \texttt{y}, \texttt{z}, and \texttt{w}). The stencil computation uses data from \texttt{float4} types to vectorize the computation. Subsequently, the generated code updates four values in global memory. In the \texttt{f4} template, the challenge is to ensure that the stencil index is accurately mapped to the current \texttt{float4} instance and its components. While manually writing vectorized code can be error-prone, using a compiler-assisted approach like the StencilPy framework significantly simplifies this process. The framework not only eases the development of vectorized code but also enhances performance through efficient vectorization of computations.

Transitioning to templates employing 2.5D blocking, the code generators must determine the stencil point index relative to the 2D plane being streamed. Next, depending on the template and its parameters (\texttt{--memory-type} and \texttt{--prefetch}), they must generate code for data allocations in various memory types, including shared memory and registers. Additionally, code is generated to set the initial values in these memory spaces before generating the streaming loop.

All 2.5D templates generate a \texttt{for} statement or \texttt{while} statement iterating from the beginning streaming index to the end. During each streaming step, code is generated to prepare data in all memory types for the stencil computations of the current plane. Then, careful code generation ensures the stencil computation is performed using stencil points from the correct locations in different memory types. Once the stencil computation is completed, the results for that plane are stored back in global memory. In the case where \texttt{--prefetch} is enabled, additional code is generated to fetch the data for the next plane, while overlapping stencil computation on a prior plane.

When preparing data for the current plane after streaming down to a new plane, the \texttt{shift} template rotates values in memory space to position them correctly for the new plane index. On the other hand, the \texttt{unroll} and \texttt{semi} templates use loop unrolling to keep existing data stationary in their memory locations while loading new data to the appropriate positions. In hand-crafted code versions, we use macro preprocessing to achieve loop unrolling. However,  StencilPy's code generators can directly generate the expanded code.

The \texttt{semi} template requires additional code generation for both the forward phase and the backward phase of the Semi-stencil algorithm. It also generates code for allocating spaces for partial results and subsequently managing these partial results according to the Semi-stencil algorithm during computations.

When \texttt{--asyncmemcpy} flag is on, and the compute capability in CUDAIR satisfies, \texttt{cudagen} produces memory copy code using \texttt{\_\_pipeline\_memcpy\_async}. Additionally, \texttt{\_\_pipeline\_commit} and \texttt{\_\_pipeline\_wait\_prior} are also properly generated.

Lastly, correct device synchronization and thread synchronization are essential, and the code generators manage these processes appropriately.

\subsection{STX Code Generation (stxgen)}

\texttt{stxgen} first needs to decompose the data domain into smaller data regions. The size of a region is subject to the hardware-specific quota of \texttt{TCDM\_SIZE}, which is the size of the memory on an STX cluster. Subsequently, for each of these regions, \texttt{stxgen} generates code to request an STX cluster and offload the respective data region to it.

Next, it generates code for allocating data arrays in the TCDM memory of the STX cluster and copies data from the host to the cluster.

Then, for the \texttt{cube} template, the code generation introduces \verb|#pragma omp target| to the entire stencil kernel and a nested \verb|#pragma stx worksharing(interleave)| to the outer loop of the data region. The first pragma specifies the section of the code that executes on the cluster, while the latter pragma offloads the stencil computation of the entire 3D cubic data region to the STX Processing Units (SPUs) on the cluster. After generating code for the stencil computation, \texttt{stxgen} generates code to transfer data back to the host from TCDM.

The \texttt{plane} template generates code aligned with STX's plane scheme design. The plane scheme closely resembles the 2.5D blocking in GPUs, with the SPU accelerating streaming index arithmetic at the hardware level. It incorporates an asynchronous DMA engine, enabling concurrent, non-interfering data transfers and computations similar to the asynchronous memory copy in the latest NVIDIA GPUs.

For \texttt{plane}, the code generation adds \verb|#pragma omp parallel num_threads(2)| and checks the OpenMP thread ID.
STX's OpenMP extension reserves ID $0$ for logic running on the host and ID $1$ for code offloading to the STX cluster. Thus, \texttt{stxgen} produces code that checks the OpenMP thread ID and generates code for host and device, respectively.

For the host thread, \texttt{stxgen} generates code to transfer the stencil results from the previous streaming step stored in TCDM back to host memory. Additionally, the generated code for the host prefetches the data for the upcoming streaming plane and stores it in the buffer.

For the cluster thread, the generated code employs both \verb|#pragma omp target| and a nested \verb|#pragma stx worksharing(interleave)| to the stencil computations for the current plane. The generated code adheres to the plane scheme design, facilitating the overlap of data transfer with stencil computations.

Finally, \texttt{stxgen} generates the code that adds \verb|#pragma omp barrier| to the end of each streaming step, ensuring synchronization among all SPUs and data transfers.

Next, the \texttt{semi} template is based on \texttt{plane} template that builds on top of the concept of the plane scheme. However, during stencil computation, instead of the conventional stencil approaches, the Semi-stencil algorithm is applied. Additionally, when generating the code for data array allocations, additional spaces are required in TCDM for partial results. Note that data transfer between the host and the device is unnecessary for partial results since they are exclusively used in stencil computations and can remain in TCDM. During the generation of code for the cluster thread, the template incorporates the Semi-stencil algorithm into the streaming dimension. It conducts a forward computation across the entire streaming plane, storing partial results, followed by a backward computation using these partial results.
The final results are stored in TCDM, and the host thread can copy them from TCDM to the host memory in the coming streaming iteration.

Finally, following the offload section, all templates generate code to release the memory allocated on the requested STX clusters.

\subsection{CSL Code Generation (cslgen)}

The dataflow architecture makes CSL code generation challenging. Generating code involves controlling router configurations, managing data communications, vectorizing the stencil computation, defining the state machine and transitions, and structuring the code layout while exposing symbols.

\paragraph{Controlling Router Configurations}

First, \texttt{cslgen} generates code defining a routing plan that involves a sequence of switch positions for the router in a PE. Route designs are typically straightforward, involving actions such as receiving from one direction, sending to the opposite direction, and utilizing \texttt{Ramp} link to communicate with the enclosing PE. The route design for each of the four directions is independent and could vary. For PEs at the edge of the PE grid, route design can be more intricate, requiring careful handling of cases when a PE may only send or receive in certain directions.

In addition to route design, \texttt{cslgen} must generate code instructing routers to switch their positions, enabling data to flow through and to its destination. This is accomplished using a CSL construct known as control wavelet. Therefore, \texttt{cslgen} produces code for control wavelets and generates code to dynamically trigger these events, facilitating the switch of router positions.

\paragraph{Managing Data Communications}
Having generated router code, \texttt{cslgen} proceeds with code generation for data communications utilizing these router configurations.

Figure~\ref{figure:stencilpy-csl-stencil-point-index-pattern} shows how DFIR represents data communication using stencil point index patterns.
Following the application of Algorithms~\ref{algo:stencilpy-offset-index-to-pattern-id} and \ref{algo:stencilpy-pattern-id-dependency}, DFIR incorporates all stencil point index patterns and their dependencies for the current stencil computation.
Algorithm~\ref{algo:stencilpy-pattern-id-to-data-communication} depicts how to convert an array of stencil point index patterns with sorted dependency to data communications.

\begin{algorithm*}[t]
    \KwIn{
    $\mathbf{patterns}$: an array of stencil point index patterns computed and sorted using Algorithms~\ref{algo:stencilpy-offset-index-to-pattern-id} and \ref{algo:stencilpy-pattern-id-dependency}, respectively}

    \SetKwFunction{FDep}{Send}
    \SetKwProg{Fn}{Function}{:}{}
    \Fn{\FDep{$d, i, j$}}{
        $d' \gets $ \Switch{d}{
            \lCase{E}{\KwRet{West}; \textbf{case} $N$ \textbf{do} \KwRet{South}}
            \lCase{W}{\KwRet{East}; \textbf{case} $S$ \textbf{do} \KwRet{North}}
        }
        \Switch{(i, j)}{
            \lCase{$(1, 0)$}{Send $0$ to $d'$}
            \Case{$(1, j)$ where $j \neq 0$}{
                $sourceId \gets d \oplus j \oplus 0$ \tcc*[r]{$\oplus$ denotes string concatenation}
                $d'' \gets $ \Switch{d}{
                    \lCase{E}{\KwRet{North}; \textbf{case} $N$ \textbf{do} \KwRet{West}}
                    \lCase{W}{\KwRet{South}; \textbf{case} $S$ \textbf{do} \KwRet{East}}
                }
                Send $sourceId$ to $d''$\;
            }
            \Other{
                $sourceId = d \oplus (i - 1) \oplus j$\;
                Send $sourceId$ to $d'$\;
            }
        }
    }
    \textbf{end}

    \ForEach{id \textbf{in} patterns}{
        $d \cdot i \cdot j \gets id $ \tcc*[r]{extracts d, i, j from id}
        Send$(d, i, j)$\;
        $d^o \gets $ \Switch{d}{
            \lCase{E}{\KwRet{East}; \textbf{case} $N$ \textbf{do} \KwRet{North}}
            \lCase{W}{\KwRet{West}; \textbf{case} $S$ \textbf{do} \KwRet{South}}
        }
        Receive from $d^o$ into $id$\;
    }

    \caption{Transforming sorted stencil point index patterns to data communications.}
    \label{algo:stencilpy-pattern-id-to-data-communication}
\end{algorithm*}

For each pattern, \texttt{cslgen} generates code choreographing the data transfer. Certain data transfers may involve multiple steps that involve intermediate routers, and these are handled accordingly by state management. Additionally, some data transfers depend on the completion of communication for other stencil point index patterns, and these dependencies are also addressed in the state management.

For a 3D box-shaped stencil with radius of two  as depicted in Figure~\ref{figure:stencilpy-box3d2r-example}, upon the application of Algorithm~\ref{algo:stencilpy-pattern-id-to-data-communication}, Table~\ref{table:stencilpy-data-common-3d2r} illustrates the
data communication actions involved in each iteration.

\begin{table*}[t]
\centering
\begin{tabular}{|c|cc|cc|cc|cc|}
\hline
        & \multicolumn{2}{c|}{N}                                                                                                                                      & \multicolumn{2}{c|}{E}                                                                                                                                      & \multicolumn{2}{c|}{S}                                                                                                                                      & \multicolumn{2}{c|}{W}                                                                                                                                      \\ \hline
        & \multicolumn{1}{c|}{\begin{tabular}[c]{@{}c@{}}Send\\ Action\end{tabular}}        & \begin{tabular}[c]{@{}c@{}}Receive\\ Action\end{tabular}                & \multicolumn{1}{c|}{\begin{tabular}[c]{@{}c@{}}Send\\ Action\end{tabular}}       & \begin{tabular}[c]{@{}c@{}}Receive\\ Action\end{tabular}                 & \multicolumn{1}{c|}{\begin{tabular}[c]{@{}c@{}}Send\\ Action\end{tabular}}        & \begin{tabular}[c]{@{}c@{}}Receive\\ Action\end{tabular}                & \multicolumn{1}{c|}{\begin{tabular}[c]{@{}c@{}}Send\\ Action\end{tabular}}         & \begin{tabular}[c]{@{}c@{}}Receive\\ Action\end{tabular}               \\ \hline
Step 1 & \multicolumn{1}{c|}{\begin{tabular}[c]{@{}c@{}}Send 0 \\ to South\end{tabular}}   & \begin{tabular}[c]{@{}c@{}}Receive\\ From North\\ into N10\end{tabular} & \multicolumn{1}{c|}{\begin{tabular}[c]{@{}c@{}}Send 0\\ to West\end{tabular}}    & \begin{tabular}[c]{@{}c@{}}Receive \\ from East \\ into E10\end{tabular} & \multicolumn{1}{c|}{\begin{tabular}[c]{@{}c@{}}Send 0 \\ to North\end{tabular}}   & \begin{tabular}[c]{@{}c@{}}Receive\\ from South\\ into S10\end{tabular} & \multicolumn{1}{c|}{\begin{tabular}[c]{@{}c@{}}Send 0\\ into East\end{tabular}}    & \begin{tabular}[c]{@{}c@{}}Receive\\ from West\\ into W10\end{tabular} \\ \hline
Step 2 & \multicolumn{1}{c|}{\begin{tabular}[c]{@{}c@{}}Send N10 \\ to South\end{tabular}} & \begin{tabular}[c]{@{}c@{}}Receive\\ From North\\ into N20\end{tabular} & \multicolumn{1}{c|}{\begin{tabular}[c]{@{}c@{}}Send E10\\ to West\end{tabular}}  & \begin{tabular}[c]{@{}c@{}}Receive \\ from East \\ into E20\end{tabular} & \multicolumn{1}{c|}{\begin{tabular}[c]{@{}c@{}}Send S10 \\ to North\end{tabular}} & \begin{tabular}[c]{@{}c@{}}Receive\\ from South\\ into S20\end{tabular} & \multicolumn{1}{c|}{\begin{tabular}[c]{@{}c@{}}Send W10\\ into East\end{tabular}}  & \begin{tabular}[c]{@{}c@{}}Receive\\ from West\\ into W20\end{tabular} \\ \hline
Step 3 & \multicolumn{1}{c|}{\begin{tabular}[c]{@{}c@{}}Send N10 \\ to East\end{tabular}}  & \begin{tabular}[c]{@{}c@{}}Receive\\ From North\\ into N11\end{tabular} & \multicolumn{1}{c|}{\begin{tabular}[c]{@{}c@{}}Send E10\\ to South\end{tabular}} & \begin{tabular}[c]{@{}c@{}}Receive \\ from East \\ into E11\end{tabular} & \multicolumn{1}{c|}{\begin{tabular}[c]{@{}c@{}}Send S10 \\ to West\end{tabular}}  & \begin{tabular}[c]{@{}c@{}}Receive\\ from South\\ into S11\end{tabular} & \multicolumn{1}{c|}{\begin{tabular}[c]{@{}c@{}}Send W10\\ into North\end{tabular}} & \begin{tabular}[c]{@{}c@{}}Receive\\ from West\\ into W11\end{tabular} \\ \hline
Step 4 & \multicolumn{1}{c|}{\begin{tabular}[c]{@{}c@{}}Send N11 \\ to South\end{tabular}} & \begin{tabular}[c]{@{}c@{}}Receive\\ From North\\ into N21\end{tabular} & \multicolumn{1}{c|}{\begin{tabular}[c]{@{}c@{}}Send E11\\ to West\end{tabular}}  & \begin{tabular}[c]{@{}c@{}}Receive \\ from East \\ into E21\end{tabular} & \multicolumn{1}{c|}{\begin{tabular}[c]{@{}c@{}}Send S11 \\ to North\end{tabular}} & \begin{tabular}[c]{@{}c@{}}Receive\\ from South\\ into S21\end{tabular} & \multicolumn{1}{c|}{\begin{tabular}[c]{@{}c@{}}Send W11\\ into East\end{tabular}}  & \begin{tabular}[c]{@{}c@{}}Receive\\ from West\\ into W21\end{tabular} \\ \hline
Step 5 & \multicolumn{1}{c|}{\begin{tabular}[c]{@{}c@{}}Send N20 \\ to East\end{tabular}}  & \begin{tabular}[c]{@{}c@{}}Receive\\ From North\\ into N12\end{tabular} & \multicolumn{1}{c|}{\begin{tabular}[c]{@{}c@{}}Send E20\\ to South\end{tabular}} & \begin{tabular}[c]{@{}c@{}}Receive \\ from East \\ into E12\end{tabular} & \multicolumn{1}{c|}{\begin{tabular}[c]{@{}c@{}}Send S20\\ to West\end{tabular}}   & \begin{tabular}[c]{@{}c@{}}Receive\\ from South\\ into S12\end{tabular} & \multicolumn{1}{c|}{\begin{tabular}[c]{@{}c@{}}Send W20\\ into North\end{tabular}} & \begin{tabular}[c]{@{}c@{}}Receive\\ from West\\ into W12\end{tabular} \\ \hline
Step 6 & \multicolumn{1}{c|}{\begin{tabular}[c]{@{}c@{}}Send N12 \\ to South\end{tabular}} & \begin{tabular}[c]{@{}c@{}}Receive\\ From North\\ into N22\end{tabular} & \multicolumn{1}{c|}{\begin{tabular}[c]{@{}c@{}}Send E12\\ to West\end{tabular}}  & \begin{tabular}[c]{@{}c@{}}Receive \\ from East \\ into E22\end{tabular} & \multicolumn{1}{c|}{\begin{tabular}[c]{@{}c@{}}Send S12 \\ to North\end{tabular}} & \begin{tabular}[c]{@{}c@{}}Receive\\ from South\\ into S22\end{tabular} & \multicolumn{1}{c|}{\begin{tabular}[c]{@{}c@{}}Send W12\\ into East\end{tabular}}  & \begin{tabular}[c]{@{}c@{}}Receive\\ from West\\ into W22\end{tabular} \\ \hline
\end{tabular}
\caption{Data communication actions involved in each iteration for a 3D box-shaped stencil with radius of two  (Figure~\ref{figure:stencilpy-box3d2r-example}).}
\label{table:stencilpy-data-common-3d2r}
\end{table*}

\paragraph{Vectorizing the Stencil Computation}

To accelerate stencil computation on a Cerebras system, vectorization of computations is essential. This is achieved by using Data Structure Descriptors (DSDs) and \texttt{builtins} provided by the Cerebras SDK. A DSD is a compact representation of a set of data elements, which may be non-contiguous. A \texttt{builtin} performs a bulk operation on a DSD's elements, leveraging a single hardware instruction to vectorize computations. The Cerebras SDK bundles \texttt{builtins} supporting various numeric operators, including addition, subtraction, multiplication, multi-adds, and logical operations.

Based on the stencil computation, \texttt{cslgen} produces DSDs describing the memory access for the stencil grid involved in the computation. Subsequently, it generates numeric operations using the provided \texttt{builtins}. Given that DFIR already represents the computation in Static Single Assignment (SSA) form, \texttt{cslgen} simply needs to map operations to  \texttt{builtins}.

\paragraph{Defining the State Machine and Transitions}

The state machine serves as the glue for all subcomponents, tracking states and coordinating their transitions. It manages the preparation of sending data, the actual data transfer, the destination for received data, the arrival of all data required for stencil computation, the execution of stencil computation, the tracking of the current time iteration, and the termination of execution at the end of the time iterations.
\texttt{cslgen} also generates and manages states for the whole program execution, including its setup and teardown.

We show the generated states and illustrate state transitions using an example of a 3D box-shaped stencil with a radius of two, as depicted in Figure~\ref{figure:stencilpy-box3d2r-example}. This example assumes a straightforward stencil update with all coefficients set to one in each iteration. We iterate the stencil updates $1000$ times for demonstration purposes.

\begin{lstlisting}[float,floatplacement=t,language=Python, basicstyle=\linespread{1.2}\ttfamily\scriptsize, label={lst:stencilpy-box3d2r-stencil-sample-code}, caption=Example code for iterating a simple update for a 3D box-shaped stencil with a radius of two as in Figure~\ref{figure:stencilpy-box3d2r-example}.]
@st.kernel
def kernel_box3d2r(u: st.grid, v: st.grid):
    v.at(0, 0, 0).set( # a box3d2r stencil update assuming coefficients of ones
        u.at(-2, 0, 0) + u.at(-2, -2, -2) + u.at(-2, -2, -1)
        + u.at(-2, -2, 0) + u.at(-2, -2, 1) + u.at(-2, -2, 2)
        + ...  # omitted for brevity
    )

@st.target
def target_box3d2r(u: st.grid, v: st.grid):
    for _t in range(1000):
        st.map(e=u.shape)(kernel_box3d2r)(u, v)
        (v, u) = (u, v)
\end{lstlisting}

First, \texttt{cslgen} generates three states: \texttt{STATE\_SETUP}, \texttt{STATE\_TEARDOWN}, and \texttt{STATE\_EXIT}.
\texttt{STATE\_SETUP} prepares the execution preparation, during which the timer is started.
\texttt{STATE\_TEARDOWN} finishes the execution, stopping the timer. Moreover, \texttt{STATE\_TEARDOWN} prepares the buffers for the final results, including the stencil data arrays and the profiling data, allowing their transfer back to the host.
\texttt{STATE\_EXIT} places the fabric in its final state and signals the host to start receiving the final results.

DFIR contains information regarding the dependencies among stencil point index patterns. Consequently, \texttt{cslgen} generates code defining the states necessary for these stencil point index patterns and produces code facilitating transitions among these states based on their dependencies. For each index pattern, \texttt{cslgen} generates two states, \texttt{STATE\_PREP\_TRANS\_$id$} and \texttt{STATE\_TRANS\_$id$}, for preparing and initiating the data transfers, respectively.
\texttt{STATE\_PREP\_TRANS\_$id$} loads the data being transferred to a dedicated data buffer and makes it available to the router. On the other hand, \texttt{STATE\_TRANS\_$id$} activates the router to enable the data flow into the router links. Once all data has been piped into the links, it controls the router switches to start receiving data, which is then stored into the appropriate data buffer. Algorithm~\ref{algo:stencilpy-pattern-id-to-data-communication} outlines this transformation.
For the example of a 3D box-shaped stencil, \texttt{cslgen} generates the following states:
\begin{itemize}
    \item {STATE\_PREP\_TRANS\_$10$},
    \item {STATE\_TRANS\_$10$},
    \item {STATE\_PREP\_TRANS\_$20$},
    \item {STATE\_TRANS\_$20$},
    \item {STATE\_PREP\_TRANS\_$11$},
    \item {STATE\_TRANS\_$11$},
    \item {STATE\_PREP\_TRANS\_$21$},
    \item {STATE\_TRANS\_$21$},
    \item {STATE\_PREP\_TRANS\_$12$},
    \item {STATE\_TRANS\_$12$},
    \item {STATE\_PREP\_TRANS\_$22$}, and
    \item {STATE\_TRANS\_$22$}.
\end{itemize}

Besides, \texttt{cslgen} generates two more states, \texttt{STATE\_UPDATE\_STENCIL} and \texttt{STATE\_ITERATION\_CHECK}, to execute the actual stencil computation and perform time iteration checking, respectively. Control statements such as ternary operators and if-else statements are also converted into states, transforming conditionals into explicit state transitions.
For instance, in the stencil computation depicted in Listing~\ref{lst:stencilpy-box3d2r-stencil-sample-code}, which iterates $1000$ times, \texttt{cslgen} generates code in \texttt{STATE\_ITERATION\_CHECK} such that if the currently processed iterations are less than $1000$, it transitions to \texttt{STATE\_PREP\_TRANS\_$10$}; otherwise, it transitions to \texttt{STATE\_TEARDOWN}.

In addition to generating all the states, \texttt{cslgen} also produces code that handles state transitions. We assume the entire state machine always starts with \texttt{STATE\_SETUP}. Upon the setup step finishes, it transitions to the first state generated based on stencil point index pattern dependencies. In our example of the 3D box-shaped stencil, this is \texttt{STATE\_PREP\_TRANS\_$10$}.
Once the state machine progresses through all data communication states, the last communication state transitions to \texttt{STATE\_UPDATE\_STENCIL}. In our example, upon the completion of \texttt{STATE\_TRANS\_$22$}, it transitions to \texttt{STATE\_UPDATE\_STENCIL}. Upon finishing the stencil update for this iteration, it transitions to \texttt{STATE\_ITERATION\_CHECK}. Depending on the iteration count, it either returns to \texttt{STATE\_PREP\_TRANS\_$10$} for data communication and stencil computation for the next iteration, or it transitions to \texttt{STATE\_TEARDOWN} to wrap up the execution. Finally, it transitions to \texttt{STATE\_EXIT} to conclude the state machine transitions.

\paragraph{Structuring the Code Layout While Exposing Symbols}

Lastly, \texttt{cslgen} generates a layout file that instructs the Cerebras runtime on the distribution of the ELF binaries.
It creates code that petitions a rectangular fabric region for ELF execution. Each PE within the requested region specifies the generated CSL source file containing router configurations, data communication, stencil computations, and the state machine. It sets up the initial state of the program execution.

Additionally, the code generator produces instructions for exposing symbols, enabling the host to access pointers to either data buffers or functions. Exposing data buffers to the host facilitates the transfer of data to or from the fabric, while function pointers enable the host to invoke them, initiating the state machine and receiving callbacks upon its completion.

\section{Customizations and Extensions}\label{sec:stencilpy-cus-ext}

StencilPy's modular design supports customization and extension at various levels, including parameterized backend, derivative backend, implementing new launchers for existing backends, and introducing new templates and backends.

\subsection{Parameterized Backend}

Built-in backends are parameterized to allow configuration of framework behaviors at compile and/or run time as discussed in Section~\ref{sec:stencilpy-opt}.

\subsection{Derivative Backend}

The framework supports derivative backends, enabling the modification or minor extension of an existing backend. This concept is similar to creating a subclass in object-oriented programming, where derived backends can be inherited from an existing backend. Derived backends define overridden behaviors or introduce new behaviors, adapting them to the existing backend behavior.

For instance, in the context of the dataflow architecture, our CSL backend currently supports the Cerebras CS-2 system with behaviors tailored to that specific machine using their SDK version 1.0.0, such as allocating extra PEs used as buffers for host-to-device data communications. SDK version 1.0.0 allocates extra three, one, four, and one PEs to the north, east, south, and west boundaries, respectively. If the sizes of these extra PE allocations ever change due to machine upgrades or SDK updates, one can introduce a derived backend. This derived backend can either overwrite these configurations or set both the PE dimensions and fabric dimensions directly without significantly altering the framework.

\subsection{New Launchers}

The framework supports the addition of new launchers, enabling alterations in the way kernels are launched or supporting new hardware by reusing existing generated code while executing it with configurations dedicated to the new hardware.

For example, our SYCL backend is presently designed for data center-grade Intel discrete GPUs with dedicated on-device memory. SYCL, being a cross-platform abstraction layer, also supports GPUs from NVIDIA and AMD, CPUs, and FPGAs. If there's a requirement to reuse the generated SYCL code but launch it on these alternative platforms, one can effortlessly implement new launchers. The framework maintains the same SYCL configuration and code generation, leveraging the new launcher to execute binaries on these diverse platforms.

\subsection{New Templates and Backends}

The framework's modular design facilitates its extension through the introduction of new templates to an existing backend and the implementation of entirely new backends for future architectures. The framework supplies the HIR and all configuration parameters to a backend. Interfaces for code generators and launchers are also provided. Common IR and shared code generation utility functions are extracted into internal libraries to ease code reuse.

For those looking to implement their own backend, any of our built-in backends can serve as an example. This approach also streamlines the process and ensures consistency in extending and implementing custom backends within the framework.

\section{Evaluations}\label{sec:stencilpy-eval}

This section describes our evaluation, starting with the numerical accuracy of the generated code.
Then, we use a 3D 25-point star-shaped stencil to evaluate its runtime performance.
Next, we quantitatively demonstrate the developer productivity boost accomplished by using the framework.

Our evaluation covers all existing backends and templates that are built-in to the current version of the framework.

\subsection{Numerical Accuracy}

The framework can generate backend code for various orders and of different shapes, so to evaluate the numerical accuracy of the generated code on various backends, we assembled a suite of kernels commonly used in stencil computations based on stencils described by Matsumura et al.~\cite{matsumura_an5d_2020_artifact}.
These kernels encompass a wide range of variations in terms of stencil characteristics found in scientific applications.

Table~\ref{table:stencilpy-kernels-eval} provides an overview of these kernels and highlights the properties of interest for our evaluations.
We have included both star-shaped and box-shaped stencils, as well as commonly used stencils like those found in Jacobian matrices.
Our evaluation includes both 2D and 3D stencils, with stencil orders ranging from 1 to 4.

For the STX backend, since 2D kernels are not supported by STX toolchain, only 3D kernels are evaluated.
Moreover, although 2D stencils can be evaluated for the Cerebras backend, they represent the least efficient utilization of the hardware. Therefore, we exclude them from our evaluation as they lack practical significance.

Although we are confident that our \texttt{cslgen} generates correct code for \texttt{box3d3r} and \texttt{box3d4r} kernels, the \texttt{cslc} compiler crashes internally during frame lowering.
We have reported the issue to the vendor, but in this study, we have omitted their evaluations.

\begin{table}
\centering
\begin{tabular}{|c|c|c|r|r|}
\hline
Kernel     & \begin{tabular}[c]{@{}c@{}}Stencil \\ Shape\end{tabular} & Dimension           & \multicolumn{1}{c|}{\begin{tabular}[c]{@{}c@{}}Stencil \\ Order\end{tabular}} & \multicolumn{1}{c|}{FLOPs} \\ \hline
star2d1r   & \multirow{8}{*}{Star}                                    & \multirow{4}{*}{2D} & 1                                                                             & 9                          \\ \cline{1-1} \cline{4-5}
star2d2r   &                                                          &                     & 2                                                                             & 17                         \\ \cline{1-1} \cline{4-5}
star2d3r   &                                                          &                     & 3                                                                             & 25                         \\ \cline{1-1} \cline{4-5}
star2d4r   &                                                          &                     & 4                                                                             & 33                         \\ \cline{1-1} \cline{3-5}
star3d1r   &                                                          & \multirow{4}{*}{3D} & 1                                                                             & 13                         \\ \cline{1-1} \cline{4-5}
star3d2r   &                                                          &                     & 2                                                                             & 25                         \\ \cline{1-1} \cline{4-5}
star3d3r   &                                                          &                     & 3                                                                             & 37                         \\ \cline{1-1} \cline{4-5}
star3d4r   &                                                          &                     & 4                                                                             & 49                         \\ \hline
box2d1r    & \multirow{8}{*}{Box}                                     & \multirow{4}{*}{2D} & 1                                                                             & 17                         \\ \cline{1-1} \cline{4-5}
box2d2r    &                                                          &                     & 2                                                                             & 49                         \\ \cline{1-1} \cline{4-5}
box2d3r    &                                                          &                     & 3                                                                             & 97                         \\ \cline{1-1} \cline{4-5}
box2d4r    &                                                          &                     & 4                                                                             & 161                        \\ \cline{1-1} \cline{3-5}
box3d1r    &                                                          & \multirow{4}{*}{3D} & 1                                                                             & 53                         \\ \cline{1-1} \cline{4-5}
box3d2r    &                                                          &                     & 2                                                                             & 249                        \\ \cline{1-1} \cline{4-5}
box3d3r    &                                                          &                     & 3                                                                             & 685                        \\ \cline{1-1} \cline{4-5}
box3d4r    &                                                          &                     & 4                                                                             & 1457                       \\ \hline
j2d5pt     & Star                                                     & 2D                  & 1                                                                             & 10                         \\ \hline
j2d9pt-gol & Box                                                      & 2D                  & 1                                                                             & 18                         \\ \hline
j2d9pt     & Star                                                     & 2D                  & 2                                                                             & 18                         \\ \hline
j3d27pt    & Box                                                      & 3D                  & 1                                                                             & 54                         \\ \hline
\end{tabular}
    \caption{Kernels for numerical correctness evaluation}
    \label{table:stencilpy-kernels-eval}
\end{table}

To assess numerical accuracy, our framework processes these kernels and generates multiple code versions using different backends, templates, template customizations, and optimization parameters.
We designed the experiment to be comprehensive to cover all variants of our code generators. We generate many
 code versions for each backend.
The numbers of generated code versions for each backend are:
\begin{multicols}{3}
\begin{itemize}
    \item seq: 20
    \item omp: 200 
    \item cuda: 648
    \item hip: 348
    \item sycl: 348
    \item stx: 21
    \item csl: 7
\end{itemize}
\end{multicols}


The evaluation runs a total of 1592 code versions to check the numerical accuracy.
In our evaluation, we automate the comparison of results of kernels generated with StencilPy against a separately hand-crafted reference OpenMP implementation to verify numerical accuracy.

For the result of each stencil point, we define an error as the absolute numerical difference between a kernel generated with StencilPy and the reference OpenMP implementation.
We find the maximum error and calculate the root-mean-square deviation (RMSD) of all errors for the entire stencil data grid.
The values involved in the computation range from $10^{-4}$ to $10^{5}$.
Our max errors for these comparisons are in the range of $10^{-7}$, while RMSDs are in the magnitude of $10^{-8}$,
proving good numerical accuracy of our code generation.

\subsection{Runtime Performance and Portability}

To evaluate the StencilPy framework in terms of performance, portability, and productivity using real-world scientific applications, we present a 25-point star-shaped stencil used in the acoustic isotropic approximation of the wave equation~\cite{meng_minimod_2020} (Acoustic ISO).
Previous works by \hbox{Raut et al.}~\cite{eric_minimod_openmp}, Sai et al.~\cite{ryuichi_cpe_2021,ryuichi_pmbs_2021,ryuichi_rice_energyhpc_2023}, and Jacquellin et al.~\cite{MauricioSC22,cerebras_sdk_doc} present manually developed code versions for CPUs, GPUs, STX, and Cerebras, respectively.
In this study, we implement the Acoustic ISO stencil using the StencilPy DSL, and we use the hand-crafted versions as reference implementations.
For a fair comparison, we also rerun the evaluations on our hardware with newer software stacks.

We compare the overall {\em time-to-solution} of code generated by StencilPy with that of hand-crafted code. This includes execution time both on the host and on the accelerator, if any.
For StencilPy, time-to-solution includes the cost of parsing and code generation.

To facilitate the performance evaluation of code generated by StenncilPy, we added a built-in profiler to the framework. This profiler collects and reports performance metrics, including time for code parsing, code generation, compilation, data movements between host and device, execution on devices, etc.

\subsubsection{CPUs}
For CPU evaluations, we first present results using a MacBook Pro laptop and then discuss two data-center-level many-core CPU devices: AMD EPYC 9684X and Fujitsu A64FX.

\paragraph{Apple M1 Max}

First, we present the time measurement results on a MacBook Pro equipped with the Apple Silicon M1 Max chip. Despite being a laptop designed for consumer use and lacking data-center-level hardware features, it serves as one of our development machines during tool development.
Therefore, the performance results on Apple Silicon offer insights into the average user experience for typical scientific application developers.

The M1 Max chip employs an ARM-based architecture. The M1 Max used in our development and evaluations hosts eight performance cores with a maximum frequency of 3.2 GHz and two efficiency cores clocked at 2.06 GHz. The machine is equipped with 32 GB of RAM. The code compilation uses Apple Clang version 15.0.0, an Apple-specific Clang distribution integrated into macOS' development toolchains.

\begin{table*}[t]
    \begin{center}
\includegraphics[scale=0.72,trim={3.3cm 12.9cm 3.3cm 0},clip]{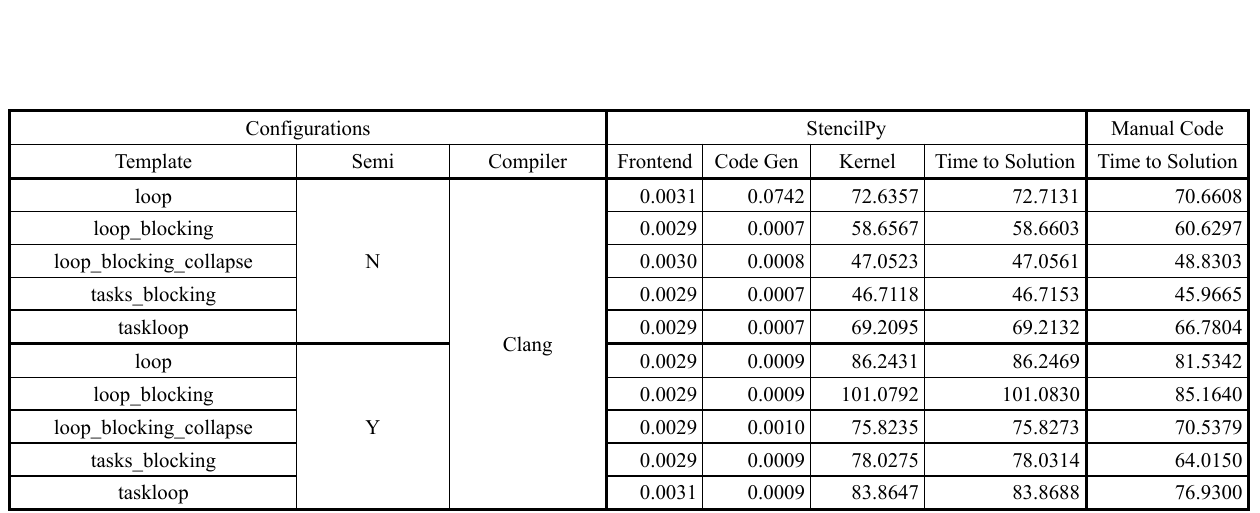}
\caption{Time measurements on a Macbook Pro laptop with Apple Silicon M1 Max chip.}
\label{table:m1max-time-measurement}
\end{center}
\end{table*}

Due to memory size constraints, we conduct runs with a grid size of $500^3$.
The evaluation runs $1000$ time iterations.
The time measurement results are detailed in Table~\ref{table:m1max-time-measurement}.
All time measurements are presented in seconds.

The overall time to solution from StencilPy-generated code is comparable to that of manual code. The overhead introduced by the framework remains relatively low, negligible when compared to the overall time to solution. Consequently, the iteration time for a developer to assess a code change with a substantial data grid size is less than two minutes, a feedback cycle deemed quite reasonable and normal in daily practices.

\paragraph{AMD GenoaX}

We conducted evaluations on a system equipped with an AMD EPYC 9684X CPU of 192 cores. The machine is configured with 750 GB of RAM. The evaluation environment supports both GCC 8.3.1 and Clang 17.0.0 compilers, so our experiments assess both compilers. To ensure fair comparisons, our runs are conducted with a grid size of $500^3$ and time iterations of $1000$.

\begin{table*}[t]
    \begin{center}
\includegraphics[scale=0.72,trim={3.3cm 6.6cm 3.3cm 0},clip]{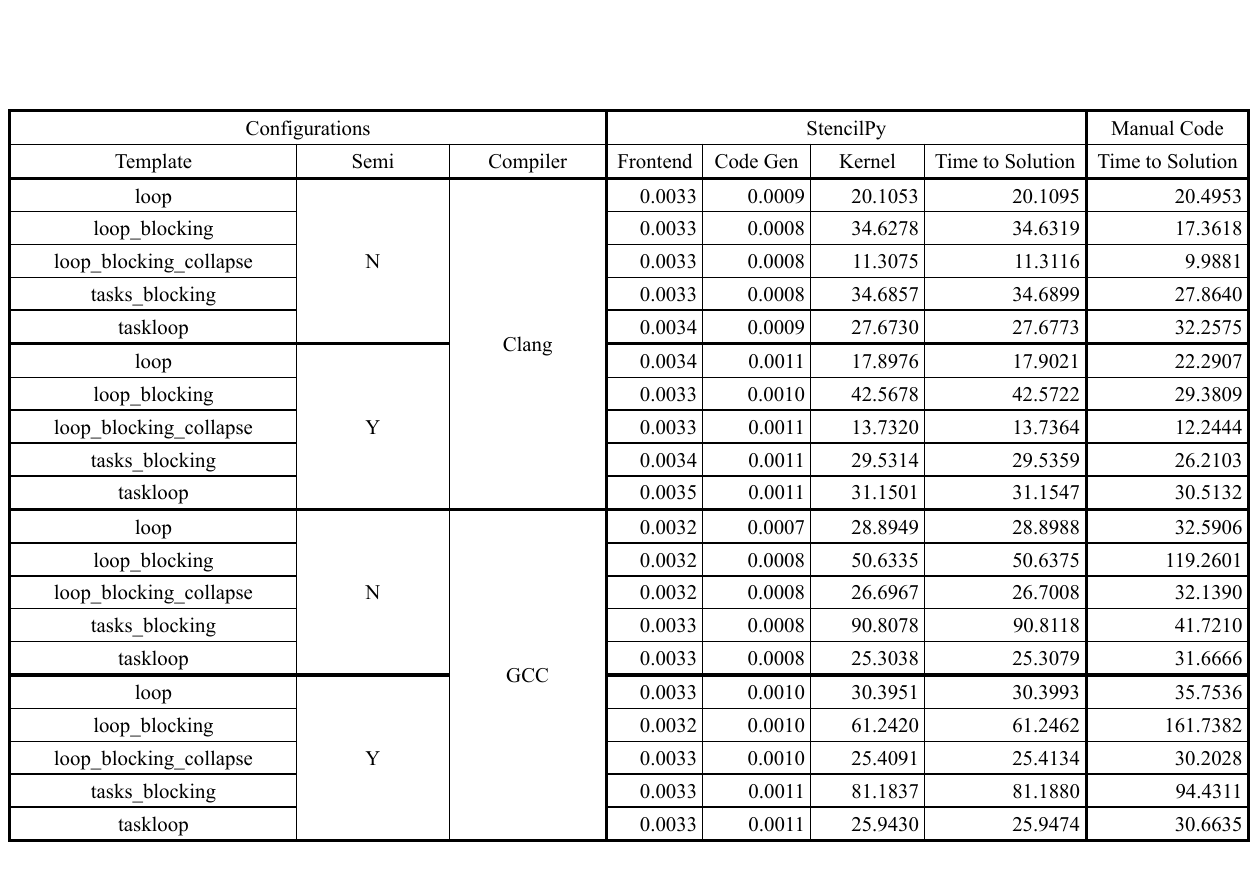}
\caption{Time measurements on an AMD GeonaX CPU}
\label{table:genoax-time-measurement}
\end{center}
\end{table*}

The results of time measurements are outlined in Table~\ref{table:genoax-time-measurement}. All time measurements are in seconds. The overall time to solution from StencilPy-generated code mostly aligns with that of manual code. Notably, in certain configurations — such as applying the taskloop template on the Clang compiler, combining the loop template with the Semi-stencil algorithm on the Clang compiler, and many configurations employing the GCC compiler — StencilPy-generated code demonstrates superior performance compared to the manual code.

The variation in performance impact among different compilers stems from the fact that OpenMP is an open standard, and each compiler has its own internal implementations. Consequently, our results reveal disparities in performance when using different compilers.
For instance, in our results, regardless of whether generated by StencilPy or manually written, the GCC compiler produces suboptimal machine code compared to the Clang compiler. Specifically, when utilizing the loop\_blocking template with the Clang compiler, regardless of whether applying Semi-stencil or not, the manual code exhibits faster execution than the StencilPy-generated code. However, when employing the same configurations with the GCC compiler, while both run slower, manual code execution demonstrates significant performance degradation and becomes slower than StencilPy-generated code.

\paragraph{Fujitsu A64FX}

We assessed the A64FX backend on a system equipped with 48 A64FX cores and 32 GB RAM. To accommodate memory size constraints, our runs were conducted with a grid size of $500^3$. Each run iterates $1000$ times. The evaluations employed FCC, a proprietary Fujitsu compiler tailored with optimizations specifically designed for the Scalable Vector Engine (SVE) integrated with the A64FX chip. FCC is provided as part of Fujitsu  Software Compiler Package V1.0L21, and Python is from anaconda3 v2021.05.

\begin{table*}[t]
    \begin{center}
\includegraphics[scale=0.72,trim={3.3cm 12.9cm 3.3cm 0},clip]{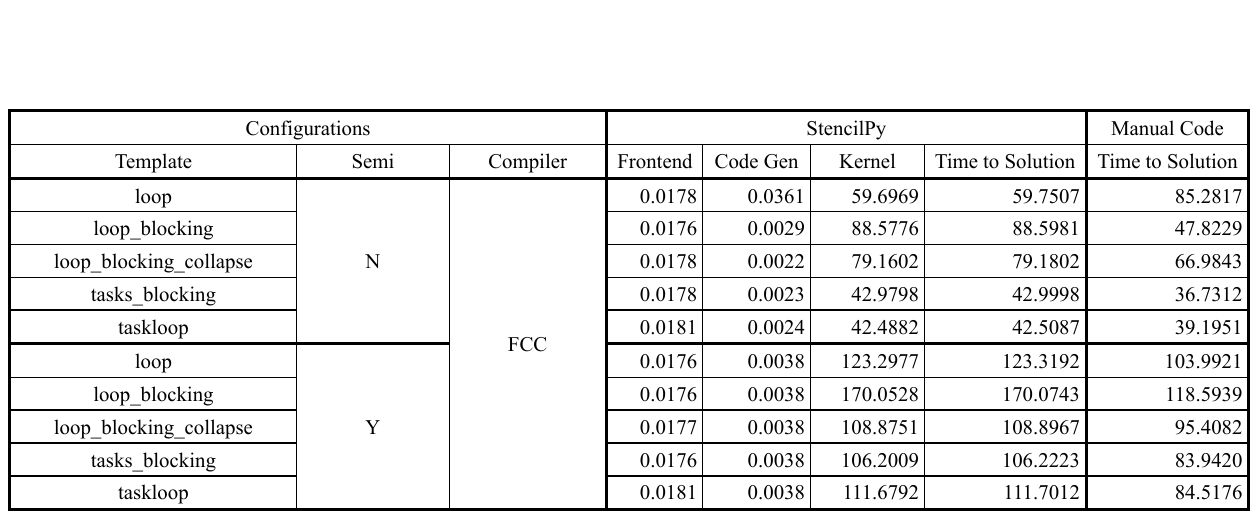}
\caption{Time measurements on A64FX.}
\label{table:A64FX-time-measurement}
\end{center}
\end{table*}

Table~\ref{table:A64FX-time-measurement} shows our framework generates code suffers a performance decline compared to the manually crafted code. Although its overhead is still deemed negligible in the context of the entire execution time, it appears to be higher than that of other backends.

We tentatively attribute the performance degradation to the machine's outdated software stack, as similar behavior is not observed on other platforms. Its Python setup may also be somewhat incompatible with Fujitsu toolchains. However, further investigations are required to gain a more confirmative understanding towards these unexpected outlier observations.

\subsubsection{GPUs}

We conducted our evaluations across multiple generations of AMD, Intel, and NVIDIA GPUs.
For NVIDIA GPUs, we evaluated on V100, A100, and two H100 GPUs. One H100 GPU is part of an NVIDIA Grace Hopper H200 Superchip and has HBM3e device memory, while the other is a PCIe Gen5 version with HBM2e device memory.
For AMD GPUs, we evaluated on AMD MI100 and MI210 GPUs.
Our evaluations also include Intel Iris P580 and Max 1100 GPUs.
This article only reports results for selected GPUs, namely NVIDIA H100 (GH200), NVIDIA H100 (PCIe), NVIDIA A100, AMD MI210, and Intel Max 1100.
Table~\ref{table:stpy-machine-spec} lists the specifications of these systems and their respective software stacks.
We refer to these systems by their GPU models.

\begin{table*}
\centering
\begin{tabular}{| c | c | c | c | c | c |}
\hline
    & H100 (GH200) & H100 (PCIe) & A100 & MI210 & Max 1100 \\
\hline
CPU & NVIDIA Grace & AMD EPYC 7313 & AMD EPYC 7402 & AMD EPYC 7402 & Intel Xeon 4410T \\
\hline
CPU Cores & 72 & 64 & 96 & 96 & 40 \\
\hline
RAM &  512 GB & 512 GB & 512 GB & 512 GB & 256 GB \\
\hline
GPU & NVIDIA H100 & NVIDIA H100 & NVIDIA A100 & AMD MI210 & Intel Max 1100 \\
\hline
Cores & 16896 & 14592   & 6912  & 6656 & 56 \\
\hline
GRAM & 95 GB & 80 GB & 40 GB & 64 GB & 48 GB \\
\hline
Platform & CUDA 12.2 & CUDA 12.0 & CUDA 12.1 & ROCm 5.5 & OneAPI 2024.0.2 \\
\hline
GPU Driver & NV 535.154.05 & NV 525.105.17 & NV 530.30.02 & ROCm 5.5 & OneAPI 2024.0.2 \\
\hline
\end{tabular}
\caption{GPU System specifications.}
\label{table:stpy-machine-spec}
\end{table*}

To evaluate performance on the H100 (GH200), H100 (PCIe), A100, MI210, and Max 1100 GPUs, we measured the execution time of $1000$ iterations.
We run different code versions with various configurations.
For each 3D template, we run with different 3D tile sizes (\verb|Dx|, \verb|Dy|, and \verb|Dz|);
For 2.5D templates, in addition to their 2D plane size (\verb|Dx| and \verb|Dy|), they differ by the memory type (\verb|Mem Type|) used for the values at the streaming dimension and by whether to prefetch coming plane (\verb|Prefetching|) and overlap the prefetching with computation.
For NVIDIA GPUs, we run with and without asynchronous memory copy (\verb|Async Mem Cp|).
In StencilPy evaluations, we report time used for frontend (\verb|Frontend|), code generation (\verb|Code Gen|), device execution (\verb|Kernel|), and total time to solution (\verb|Time to Solution|).
For executions using hand-crafted code, we report its total time to solution (\verb|Time to Solution|).
We consider \verb|Kernel| to contain the stencil and boundary condition execution times on device.
We omit lines where we lack the manual code comparison.
All times are in seconds.

\paragraph{NVIDIA H100 (GH200), H100 (PCIe), and A100}

Our evaluation executed each kernel with a grid size of ${1000^3}$ on NVIDIA GPUs.
Time measurements for 3D and 2.5D templates on the H100 (GH200) GPU are presented in Tables \ref{table:h200-3d-time-measurement} and \ref{table:h200-25d-time-measurement}, respectively.
Similarly, time measurements for 3D and 2.5D templates on the H100 (PCIe) GPU are provided in Tables \ref{table:h100-3d-time-measurement} and \ref{table:h100-25d-time-measurement}, respectively.
Corresponding time measurements for 3D and 2.5D templates on the A100 GPU are showcased in Tables \ref{table:a100-3d-time-measurement} and \ref{table:a100-25d-time-measurement}, respectively.

For 2.5D templates, Tables~\ref{table:h200-25d-time-measurement} and \ref{table:h100-25d-time-measurement} compare various 2D plane sizes on two H100 GPUs, while Table~\ref{table:a100-25d-time-measurement} emphasizes the impacts of utilizing the asynchronous memory copy on the A100 GPU.
Additionally, we present the same set of kernels for the two H100 GPUs to demonstrate the influence of high-bandwidth memory updates in the H100 (GH200) version.

\begin{table*}[t]
    \begin{center}
\includegraphics[scale=0.81,trim={4.5cm 10.5cm 3cm 0},clip]{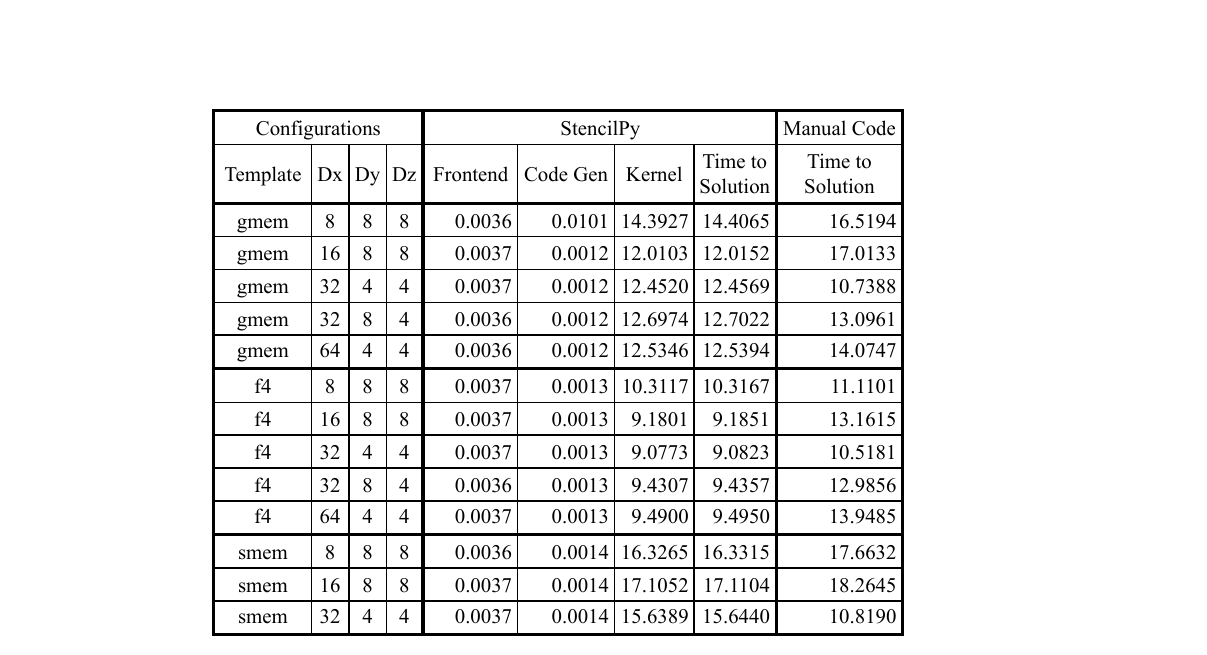}
\caption{Time measurements for 3D templates on H100 (GH200).}
\label{table:h200-3d-time-measurement}
\end{center}
\end{table*}

\begin{table*}[!htbp]
    \begin{center}
\includegraphics[scale=0.81,trim={1cm 2.1cm 1cm 0},clip]{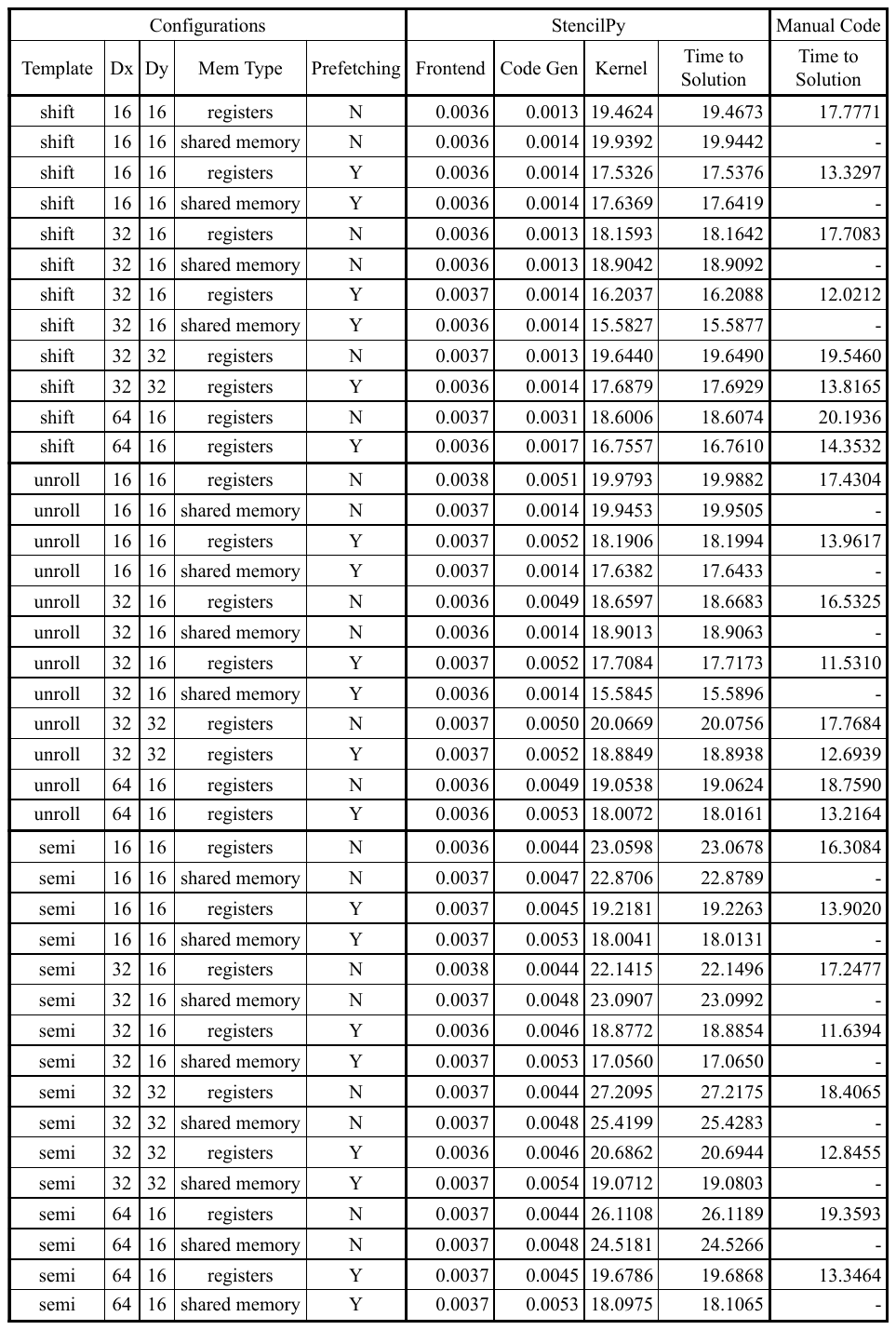}
\caption{Time measurements for 2.5D templates on H100 (GH200).}
\label{table:h200-25d-time-measurement}
\end{center}
\end{table*}

\begin{table*}[t]
    \begin{center}
\includegraphics[scale=0.81,trim={4.5cm 10.5cm 3cm 0},clip]{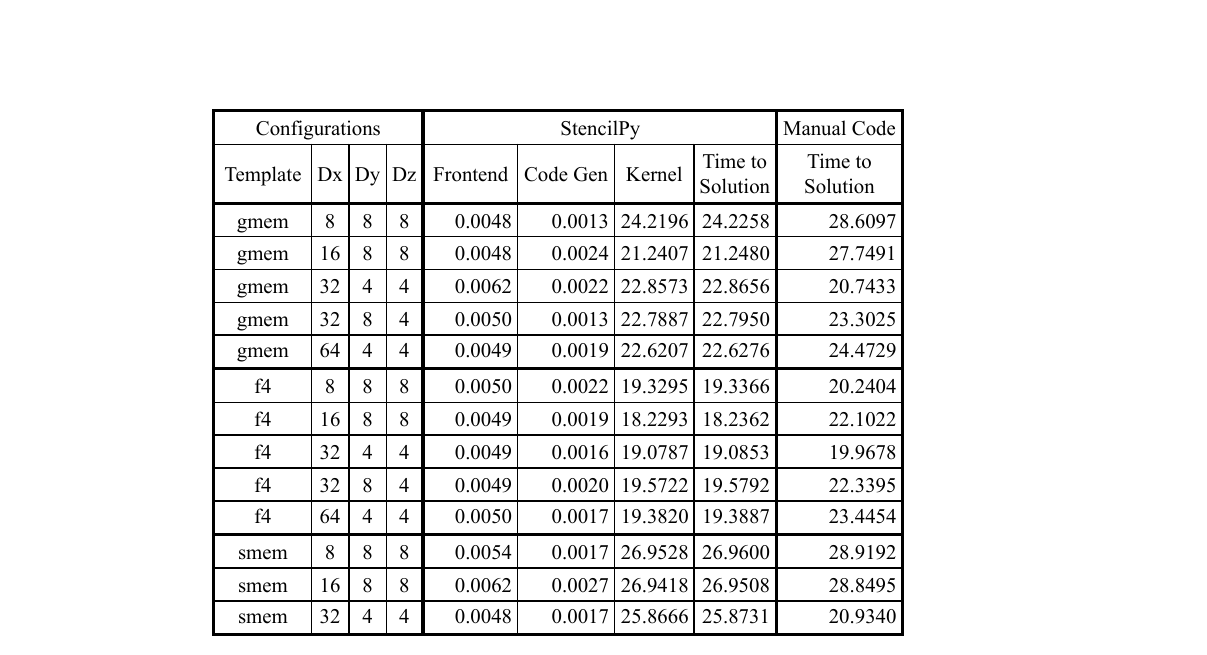}
\caption{Time measurements for 3D templates on H100 (PCIe).}
\label{table:h100-3d-time-measurement}
\end{center}
\end{table*}

\begin{table*}[!htbp]
    \begin{center}
\includegraphics[scale=0.81,trim={1cm 2.1cm 1cm 0},clip]{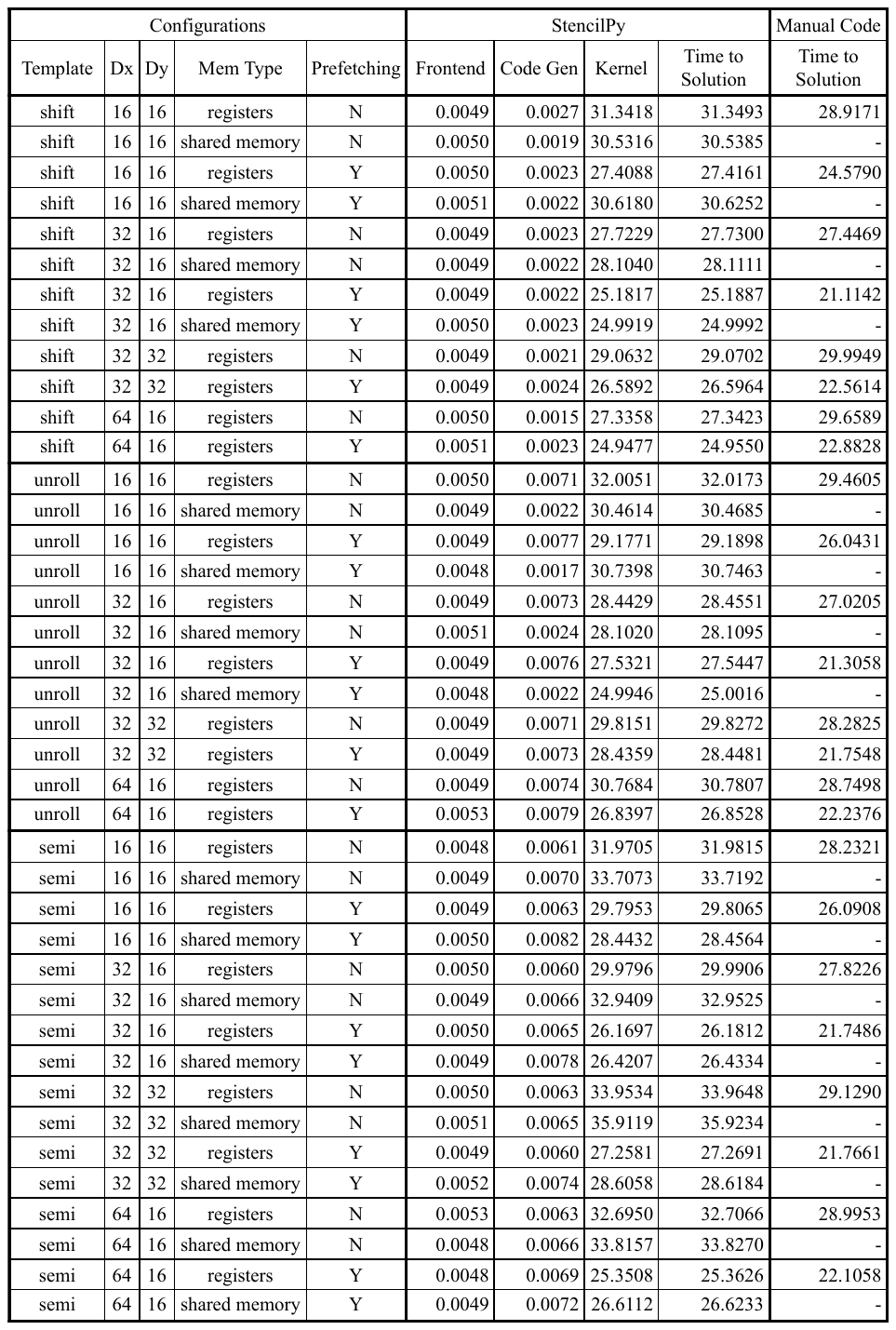}
\caption{Time measurements for 2.5D templates on H100 (PCIe).}
\label{table:h100-25d-time-measurement}
\end{center}
\end{table*}

\begin{table*}[t]
    \begin{center}
\includegraphics[scale=0.81,trim={4.5cm 10.5cm 3cm 0},clip]{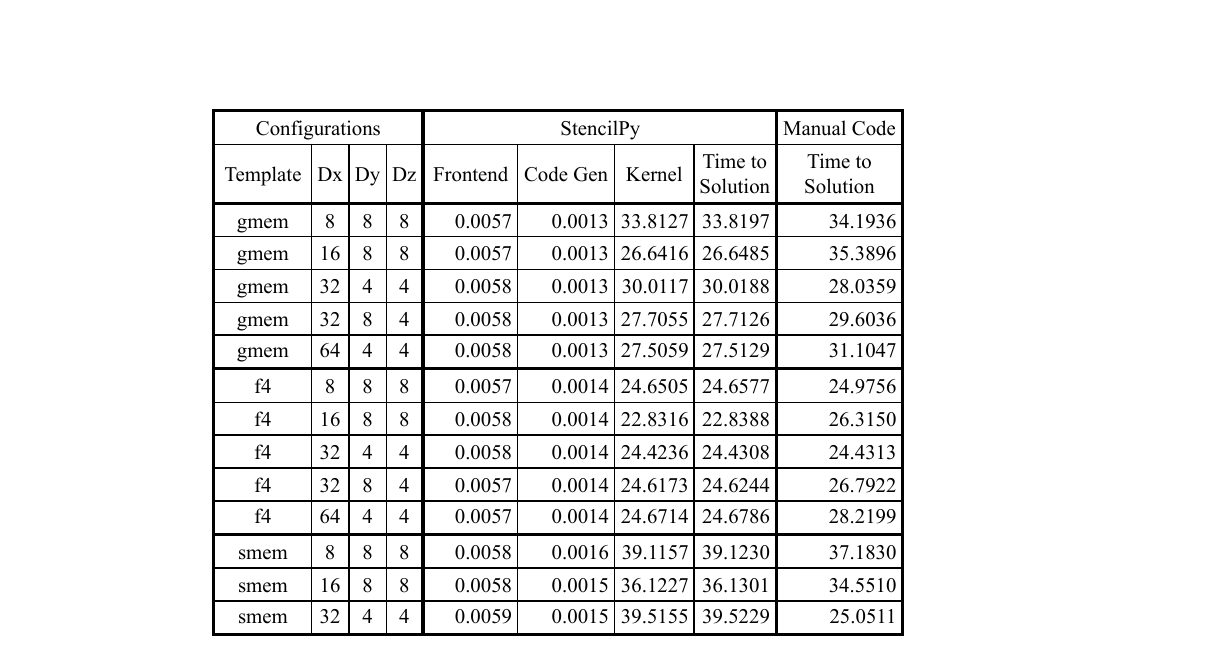}
\caption{Time measurements for 3D templates on A100.}
\label{table:a100-3d-time-measurement}
\end{center}
\end{table*}

\begin{table*}[!htbp]
    \begin{center}
\includegraphics[scale=0.81,trim={1cm 2.1cm 1cm 0},clip]{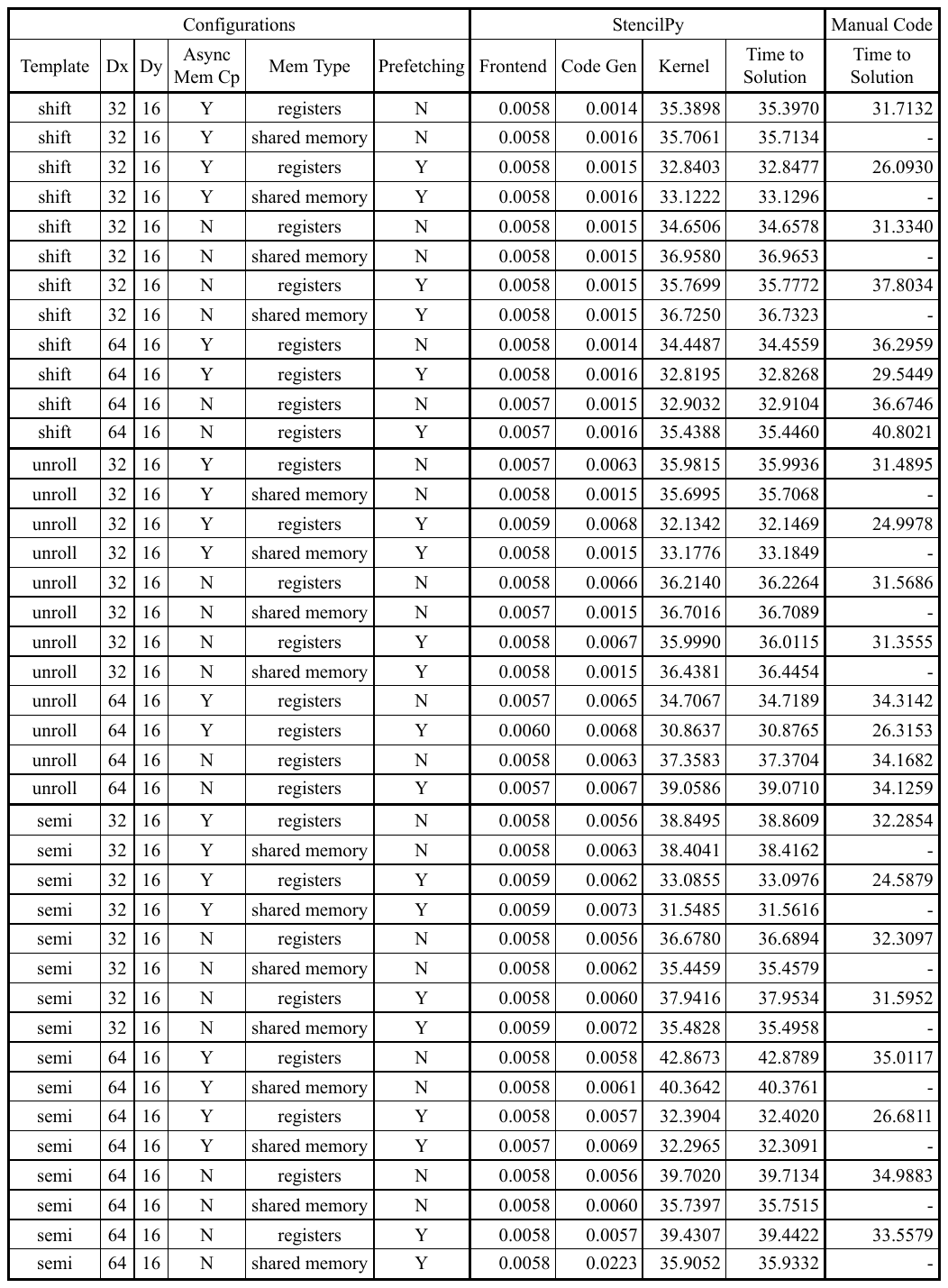}
\caption{Time measurements for 2.5D templates on A100.}
\label{table:a100-25d-time-measurement}
\end{center}
\end{table*}

Overall, our generated code performs similarly to hand-crafted code.
Across all three GPUs, the fastest \hbox{StencilPy} code version applies the \verb|f4| template.
On the H100 (GP200), the fastest kernel employing the \verb|f4| template with a block size of $32 \times 4 \times 4$, resulting in a $15.81\%$ speedup compared to the fastest manually-written CUDA kernel.
Meanwhile, the fastest \verb|f4| kernels on both the H100 (PCIe) and A100 adopt a block size of $16 \times 8 \times 8$, yielding a $9.49\%$ speedup on the H100 and a $6.92\%$ speedup on the A100 compared to the best manually-written CUDA code performances, respectively.
The tables illustrate that the overheads introduced by our framework are minimal. The time required for code analysis and generation is very small and negligible relative to the overall time to solution.

While the performance of the generated code from 3D templates is optimal, templates employing the streaming approach show room for improvement. We believe it is due to the framework generating separate conditionals that guard the initialization of a shared memory copy for each data array, even when the conditionals for each array are identical.
A potential solution involves consolidating the code into a unified set of conditionals. Additionally, another potential performance enhancement is to introduce padding between each row in the memory layout, ensuring that each row aligns with cache lines. This approach was implemented in our manual code versions but is currently missing in StencilPy's code generator. Both strategies will be explored in our future work.

\paragraph{MI210}

Tables \ref{table:mi210-3d-time-measurement} and \ref{table:mi210-25d-time-measurement} show time measurements for 3D and 2.5D templates on an MI210 GPU, respectively. These results depict the execution duration of each kernel with a grid size of ${1000^3}$.

\begin{table*}[t]
    \begin{center}
\includegraphics[scale=0.81,trim={4.5cm 9.9cm 4.5cm 0},clip]{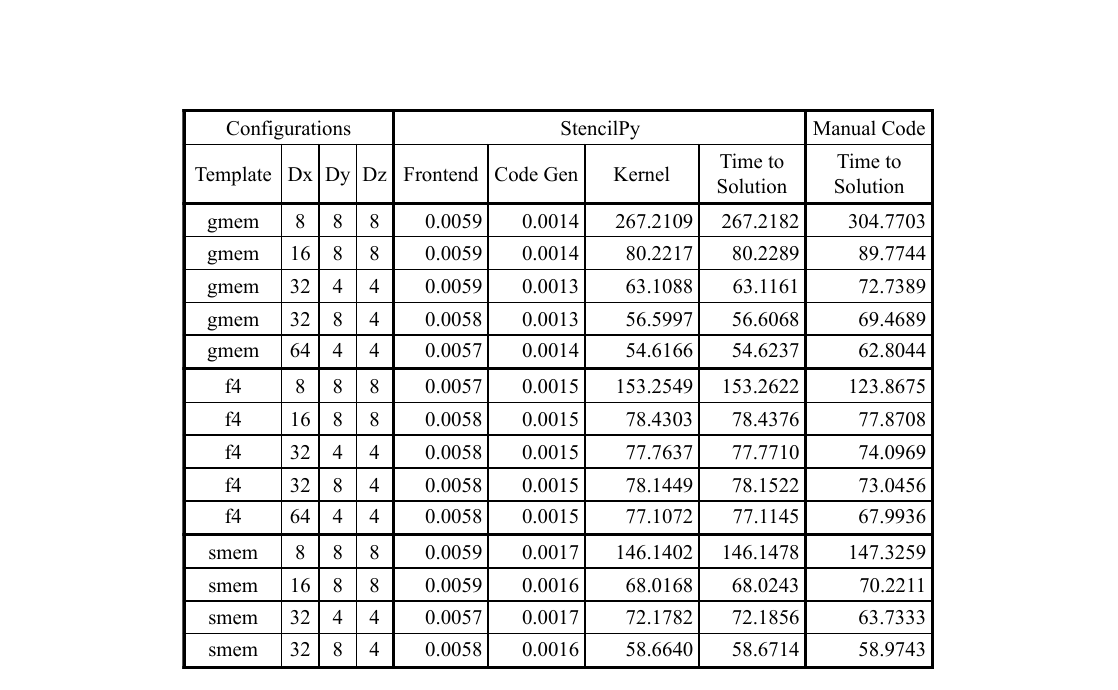}
\caption{Time measurements for 3D templates on MI210.}
\label{table:mi210-3d-time-measurement}
\end{center}
\end{table*}

\begin{table*}[!htbp]
    \begin{center}
\includegraphics[scale=0.81,trim={1cm 4.2cm 1cm 0},clip]{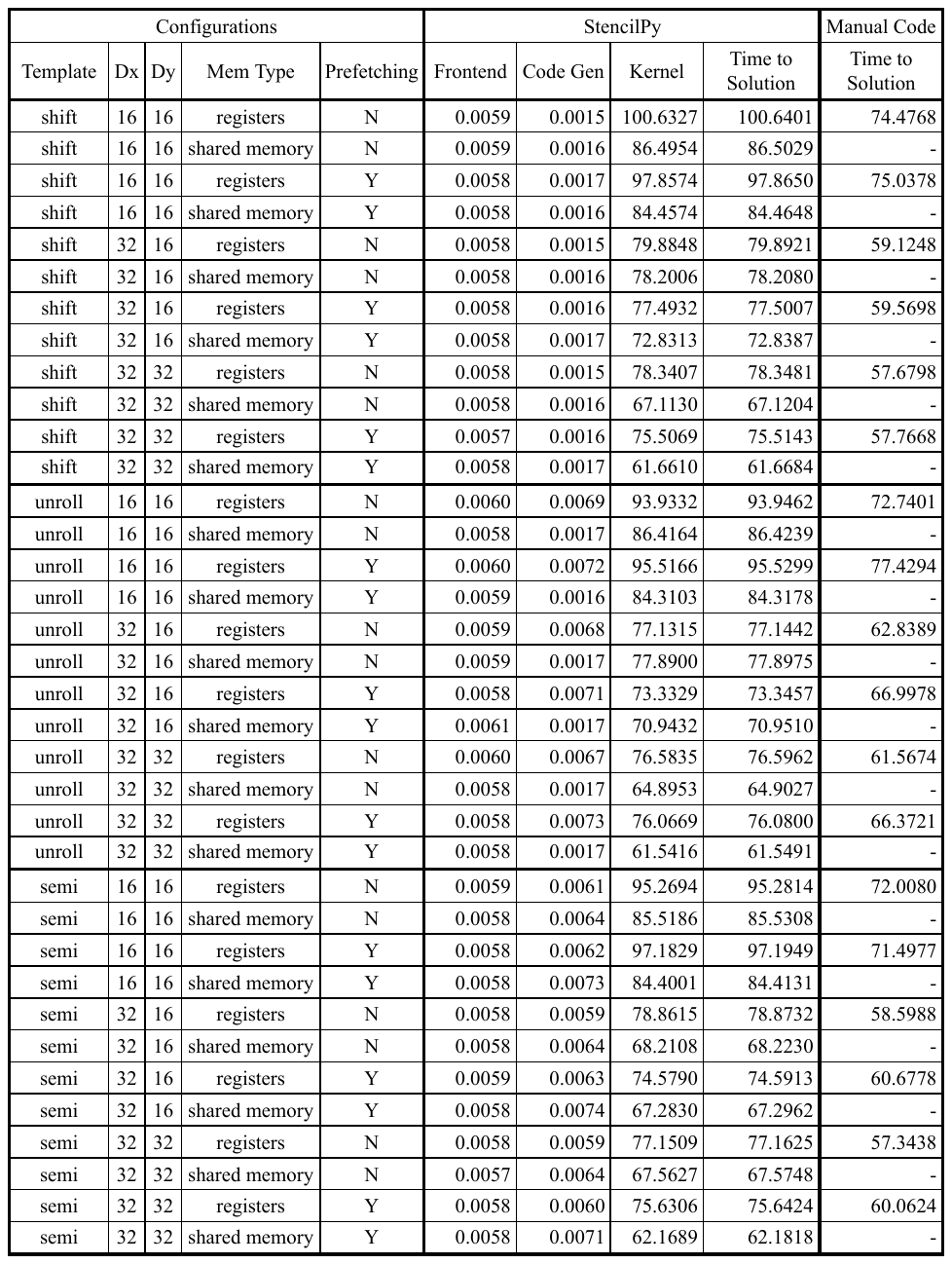}
\caption{Time measurements for 2.5D templates on MI210.}
\label{table:mi210-25d-time-measurement}
\end{center}
\end{table*}

The time measurements reveal a pattern similar to those observed with NVIDIA GPUs, highlighting that our generated code performs comparably to manually crafted code. The generated code for 3D templates exhibits speedup, while the one for 2.5D templates demonstrates degradation. We believe that the underlying issues influencing code generation for 2.5D templates remain the same.

The fastest generated code is derived from the \verb|gmem| template with a block size of $64 \times 4 \times 4$, showcasing a $4.97\%$ speedup compared to the best manually-written HIP code.

Despite improvements with newer compiler and hardware driver versions, we continue to observe suboptimal performance from the 3D blocking kernels with a shape of \texttt{8x8x8}, while the performance gaps are being minimized.
In addition to addressing the future work outlined in the NVIDIA GPUs subsection, our ongoing efforts also involve engaging with vendors to better understand the issues and explore potential mitigation strategies.

\paragraph{Max 1100}

Our evaluation executed each kernel with a grid size of ${800^3}$ on an Intel Max 1100 GPU. Time measurements for 3D and 2.5D templates on the GPU are presented in Tables \ref{table:max1100-3d-time-measurement} and \ref{table:max1100-25d-time-measurement}, respectively.

\begin{table*}[t]
    \begin{center}
\includegraphics[scale=0.81,trim={4.5cm 9.3cm 3cm 0},clip]{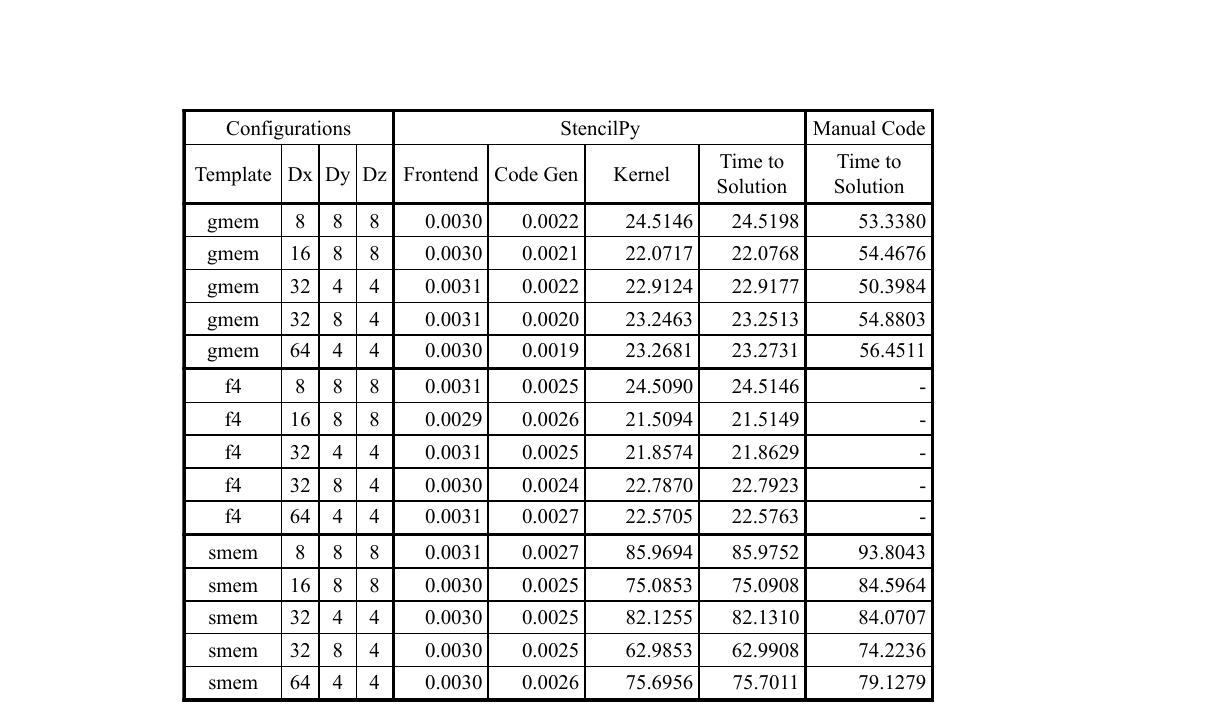}
\caption{Time measurements for 3D templates on Max 1100.}
\label{table:max1100-3d-time-measurement}
\end{center}
\end{table*}

\begin{table*}[!htbp]
    \begin{center}
\includegraphics[scale=0.81,trim={1cm 4.2cm 1cm 0},clip]{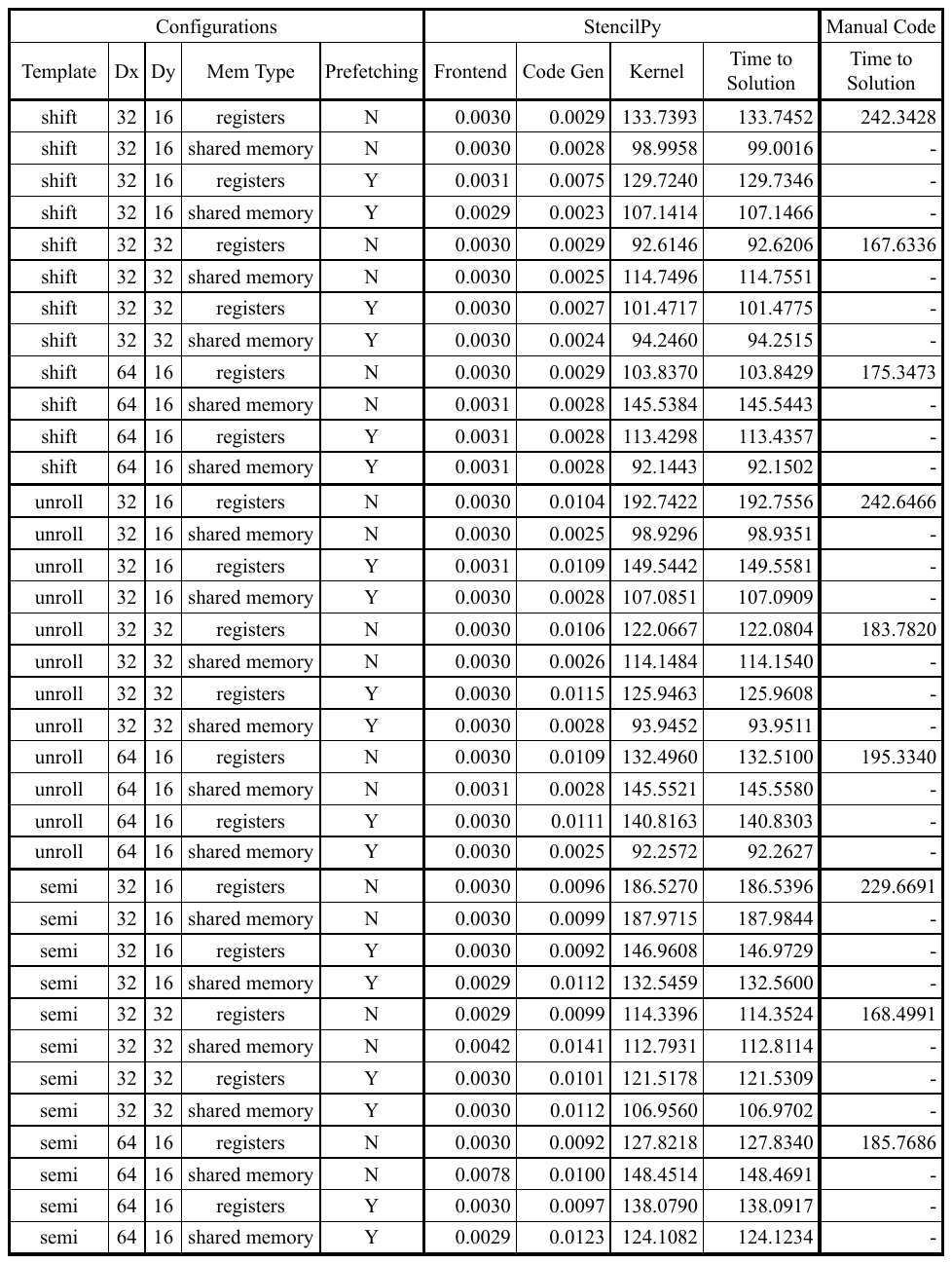}
\caption{Time measurements for 2.5D templates on Max 1100.}
\label{table:max1100-25d-time-measurement}
\end{center}
\end{table*}

We had already switched focus and started working on automated code generation when access to Intel GPU became available, so our effort spent on hand-crafted code versions was limited.
While we do not claim that our manual SYCL code is comprehensive or finely tuned, our generated code exhibits a significant speedup compared to the manual code. The most efficient StencilPy code version on the Max 1100 is from the $f4$ template with a block size of $16 \times 8 \times 8$.

It is noteworthy that loop unrolling is discouraged in the SYCL programming model~\cite{intel_oneapi_gpu_opt_guide}. However, both our generated code and manual code for the 2.5D streaming approach heavily rely on it, and thus, has a substantial impact on performance. As part of our future work, we intend to explore code generation using regular \verb|for| loops to evaluate if this approach can enhance performance.

\subsubsection{STX}

\begin{table*}[t]
    \begin{center}
\includegraphics[scale=0.87,trim={5.4cm 15cm 5.4cm 0},clip]{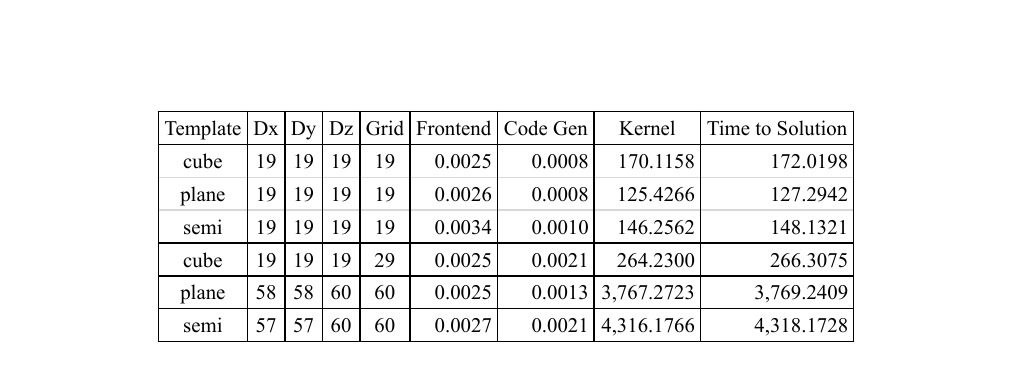}
\caption{Time measurements on STX Simulator.}
\label{table:stx-time-measurement}
\end{center}
\end{table*}

Table \ref{table:stx-time-measurement} shows time measurements for STX backend using its software simulator.
We first show three configurations of the same grid sizes (\verb|Grid|) and the same block dimensions (\verb|Dx|, \verb|Dy|, and \verb|Dz|) but use different templates.
Then, we demonstrate configurations of different block sizes that maximize the TCDM quota.
Additionally, we report the respective times used for frontend (\verb|Frontend|), code generation (\verb|Code Gen|), code compilation (\verb|Comp|), simulation time (\verb|Kernel|), and total time to solution (\verb|Time to Solution|).

Given the utilization of a software-based simulator, the kernel times are unlikely to represent the actual hardware execution time when it becomes available. Nevertheless, Table~\ref{table:stx-time-measurement} highlights that the framework demonstrates minimal overheads, with the frontend and code generation requiring negligible time duration.

It's important to note that, due to the relatively small grid sizes allowed in the software simulator, our evaluation focuses on the inner region, leaving the PML width set to zero.

\subsubsection{CSL}

Time measurements for the CSL backend are presented in Table~\ref{table:csl-time-measurement}, and for comparative purposes, additional selected backends are included in the table.
We have provided details about the evaluation environments for the other backends, while the CSL backend evaluation is carried out on a Cerebras CS-2 system running Cerebras SDK 1.0.0.

The execution time is measured by iterating the inner region of the Acoustic ISO kernel 10,000 times on each device, utilizing a grid size of ${750 \times 994 \times 300}$. This grid size maximizes both the on-device memory and computing resources of a Cerebras CS-2.
We choose a height of $300$ with the assumption that it provides enough space to concurrently store values from each neighboring cell and allocate additional temporary vector buffers needed for vectorization.
Our goal is to minimize the number of temporary variables used wherever possible.
For a fair comparison, we keep the same grid size for other backends.
The unit of measurement in Table~\ref{table:csl-time-measurement} is seconds.

\begin{table*}[t]
    \begin{center}
\begin{tabular}{|c|c|c||r|r|r|r||r|}
\hline
Backend & Template & Device & \multicolumn{1}{c|}{Frontend} & \multicolumn{1}{c|}{Code Gen} & \multicolumn{1}{c|}{Kernel} & \multicolumn{1}{c||}{Time to Solution} & \multicolumn{1}{c|}{Manual Code} \\ \hline
OMP     & loop     & Genoa  & 0.0033                        & 0.0011                        & 269.4177                     & 269.4222                               & 274.6220                 \\ \hline
CUDA    & f4       & H100   & 0.0048                        & 0.0024                        & 45.0160                     & 45.0232                               & 45.7835                 \\ \hline
HIP     & gmem     & MI210  & 0.0059                        & 0.0012                        & 138.1122                     & 138.1193                               & 150.5630                 \\ \hline
CSL     & csl      & CS2    & 0.0034                        & 0.0011                        & 0.4685                      & 0.4730                                & 0.4676                  \\ \hline
\end{tabular}
\caption{Time measurements for CSL backend and comparisons with other selected backends.}
\label{table:csl-time-measurement}
\end{center}
\end{table*}

The StencilPy-generated CSL code demonstrates performance comparable to hand-crafted code, with negligible overheads. Notably, the CSL backend exhibits a significant speedup compared to other backends.

While technically feasible, having each PE run an independent computation pattern for this type of HPC applications is still uncharted territory. Current best practices for this kind of HPC applications on a dataflow architecture necessitate uniform computation patterns on each PE. Consequently, our evaluation for CSL backend omits the PML layer and focuses solely on the inner region. The evaluation for the entire data grid is a topic for future work.

\subsection{Developer Productivity}

We evaluate developer productivity by comparing the effort required to achieve similar levels of performance. Table~\ref{table:stpy-loc} shows the number of lines of code for implementing the same Acoustic ISO kernel using StencilPy and manual implementations. StencilPy-based implementations can massively improve developer productivity with significantly less code to write.
The code is written in a logically global view, and developers do not need to know details about target architectures.
For performance tuning, developers can parameterize the framework by selecting configurations without changing the source code.

\begin{table*}[t]
\centering
\includegraphics[scale=0.9,trim={6.3cm 10.5cm 6cm 1.8cm},clip]{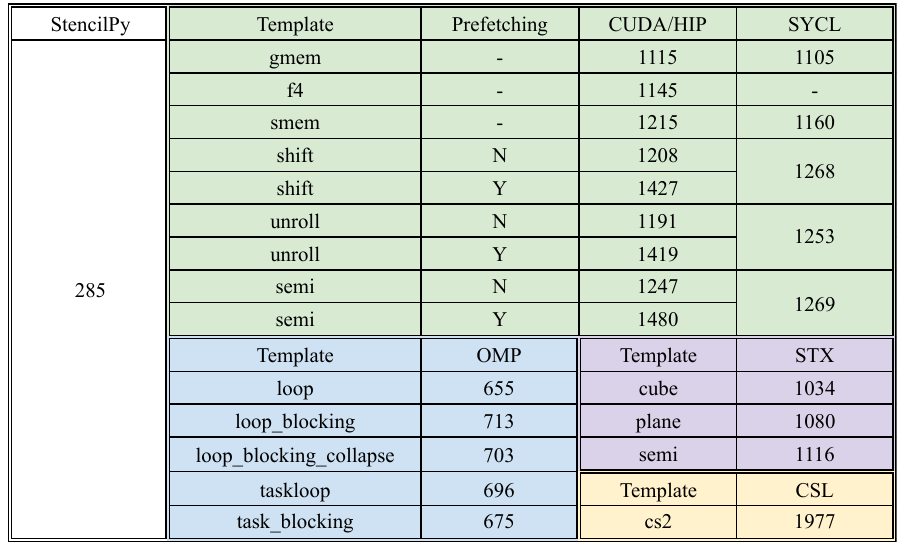}
    \caption{Lines of code for implementing Acoustic ISO using StencilPy v.s. manually crafted code.}
    \label{table:stpy-loc}
\end{table*}

\sloppy
StencilPy also improves developer productivity with a platform-agnostic DSL that can be easily ported to different hardware by reusing the same DSL code and switching to another backend.
StencilPy's modular design makes it straightforward to change the backend.

Because our DSL is embedded in Python,
developers who already know Python should feel comfortable getting started. Even for newcomers to Python, learning Python is very straightforward, especially for developers with a background in another HPC language, such as C, C++, or Fortran.

\section{Summary}\label{sec:stencilpy-summary}

We presented StencilPy, a portable framework for accelerating high-order stencils on modern CPUs, GPUs, and emerging architectures.
While still under active development, it already demonstrates promising results: it generates code for a variety of platforms with great performance, demonstrates strong performance portability spanning various hardware vendors and generations, and enhances developer productivity with a user-friendly domain-specific language.

This manuscript describes the StencilPy framework, including its multi-layered architectural design, code analysis, code generation, and execution.
It discusses the workflow of the framework and the template engine that powers the multiple backends.
It evaluates the framework from the perspectives of numerical correctness, runtime performance, performance portability, and developer productivity.
It demonstrates the approaches that can be used to accelerate high-order stencils to facilitate development, ease experiments, and reduce overheads.

\section{Future Work}\label{sec:stencilpy-future-work}

Here, we sketch possible extensions to this work.

\subsection{Optimizations to StencilPy's Generated Code}

Our work continues the performance tuning effort to further optimize the runtime performance of the generated code. Specifically, for GPU backends, our roadmap includes optimizations such as consolidating conditional checks to minimize branch divergence and using pinned memory to accelerate data transfers between the host and GPU devices. Additionally, we aim to optimize register allocations to alleviate register pressure and improve occupancy.

In the case of the SYCL programming model, where loop unrolling is not recommended, we propose generating code using a regular \verb|for| loop for the 2.5D streaming approach and evaluating the performance impacts.

For the CSL backend, we plan to introduce support for PE dimension divisions to handle larger data grids on a CS-2. In addition, we aim to reduce the footprint by carefully allocating colors and local task identifiers from a limited quota while preserving the asynchronous memory copy feature. Our proposed approach involves shifting more logic from callback tasks to state machine transitions.

Furthermore, our plans include implementing other enhancements tailored to the specifics of the programming model.

\subsection{Auto-Tuning StencilPy}

The StencilPy framework we presented in this article uses a user-guided approach, and we plan to incorporate auto-tuning capabilities.
Auto-tuning involves generating multiple code versions and dynamically selecting tuning parameters based on runtime properties and stencil characteristics.
To facilitate this, we need to define a cost model for selecting kernel candidates and introduce a feedback mechanism that allows dynamic code version switching during runtime to alternative candidates if the currently selected one fails to meet performance expectations during execution.

In addition, this capability will empower StencilPy to switch among templates and optimization strategies during runtime dynamically. As the computation progresses, the data utilized in the computation may change, such as transitioning from sparse to dense or centering around specific perturbation points before evening out. This dynamic adaptation allows the computation to efficiently utilize different templates and optimizations as the iteration unfolds, optimizing performance for long-duration executions with evolving data patterns. The built-in profiler and the Just-In-Time compilation in our framework serve as technical foundations of this extended work. Our optimized code generation and compilation require little time, making the overhead from dynamic switching negligible, especially for long-duration execution.

Furthermore, to enhance the accuracy and speed of auto-tuning, we anticipate exploring the incorporation of machine-learning techniques into the framework. This addition further refines and speeds up the auto-tuning process, ensuring optimal performance across varying stencil applications and data patterns.

\subsection{StencilPy Frontend Extensions}

We plan to introduce one additional user-facing frontend interface on top of the current domain-specific language, enabling domain application developers to express stencil computations at a higher-level abstraction. Aiming to cater to a broader user base, including geoscientists and physicists, this extension is intended to express math and physics formulas for scientific simulations that essentially use stencil computations.

\subsection{StencilPy Backend Extensions}

We plan to incorporate FPGAs as a special backend in StencilPy. StencilPy already can generate SYCL code, which inherently supports FPGAs. Our next step will involve evaluating the performance of StencilPy-generated code on FPGA devices using generated SYCL code. If the outcomes prove promising, we can consider the development of a dedicated backend tailored specifically for FPGA devices.

Furthermore, we plan to explore the feasibility of using MLIR~\cite{lattner_mlir_2020} as a virtual backend and evaluate its performance benefits. While StencilPy already supports a variety of architectures, the motivation for MLIR integration is to leverage its built-in optimization phases before generating device-specific binaries.

\subsection{StencilPy Framework Distribution}

We plan to explore Python's distribution model to facilitate support for diverse backends, including various host platforms and device machines.
Additionally, we aim to open-source the tool, increasing accessibility, collecting feedback, and fostering collaboration within the community.

\begin{acks}
This work was supported in part by a contract from \hbox{TotalEnergies EP Research \& Technology} USA, LLC.
We thank Timo Eichmann, Fran{\c c}ois Hamon, Marc Andre Heller, Mathias Jacquelin, Jens Kr\"{u}ger, Jie Meng, Kai Plociennik, and Sameer Shende for their support of this work.
We gratefully acknowledge the computing resources provided by
Joint Laboratory for System Evaluation at Argonne National Laboratory,
Oak Ridge Leadership Computing Facility,
University of Oregon,
Rice University,
and \hbox{TotalEnergies}.
\end{acks}

\bibliographystyle{abbrv}
\bibliography{main}

\end{document}